\documentclass[aps, twocolumn, showpacs, preprintnumbers, amsmath, amssymb, prb]{revtex4-2}
\usepackage{amsfonts}
\usepackage{mathrsfs}
\usepackage{amsmath}% needed for subequations
\usepackage{color}
\usepackage{natbib}
\usepackage{textcomp}
\usepackage{graphicx}
\usepackage{enumerate}
\usepackage{bm}% bold maths
\usepackage{amssymb}
\usepackage{xspace}
\usepackage{epstopdf}
\usepackage{dcolumn}% Align table columns on decimal point
\usepackage{longtable}
\usepackage{subfigure}
\usepackage{multirow}
\usepackage{makecell}
\usepackage{array}
\usepackage[colorlinks=true, letterpaper=true, pdfstartview=FitV, linkcolor=blue, citecolor=blue, urlcolor=blue]{hyperref}
\usepackage{float}
\usepackage[figuresright]{rotating}
\usepackage{longtable}
\usepackage{booktabs}
\usepackage{tabularx}
\usepackage{hyperref}
\usepackage{braket}

\makeatletter

\newcommand{\Rmnum}[1]{\expandafter\@slowromancap\romannumeral #1@}
\makeatother

\begin{document}

\title{Intrinsic Second Order Spin Current}

\author{Zhi-Fan Zhang$^{1}$}
%\affiliation{School of Physical Sciences, University of Chinese Academy of Sciences, Beijing 100049, China. }

\author{Zhen-Gang Zhu$^{2,1}$}
\email{zgzhu@ucas.ac.cn}
%\affiliation{School of Physical Sciences, University of Chinese Academy of Sciences, Beijing 100049, China. }
%\affiliation{School of Electronic, Electrical and Communication Engineering, University of Chinese Academy of Sciences, Beijing 100049, China.}
%\affiliation{CAS Center for Excellence in Topological Quantum Computation, University of Chinese Academy of Sciences, Beijing 100049, China.}

\author{Gang Su$^{3,4}$}
\email{gsu@ucas.ac.cn}
%\affiliation{
%	Kavli Institute for Theoretical Sciences, University of Chinese Academy of Sciences, Beijing 100190, China.
%}
\affiliation{
		$^{1}$ School of Physical Sciences, University of Chinese Academy of Sciences, Beijing 100049, China. \\
		$^{2}$ School of Electronic, Electrical and Communication Engineering, University of Chinese Academy of Sciences, Beijing 100049, China.\\
		$^{3}$ Kavli Institute for Theoretical Sciences, University of Chinese Academy of Sciences, Beijing 100190, China.\\
		$^{4}$ Institute of Theoretical Physics, Chinese Academy of Sciences, Beijing 100190, China.
	}
\date{\today}

\begin{abstract}
In recent years, nonlinear Hall effect has attracted great attention with three different terms contributed by Drude effect, Berry curvature dipole and Berry connection polarizability. In this work, we theoretically predict an intrinsic second order spin current induced by spin-dependent Berry curvature polarizability based on time-independent perturbation theory. We show other two second order spin conductivities contributed by the group velocity and  spin-dependent Berry curvature dipole.
%\textcolor{red}{
A two-dimensional Rashba-Dresselhaus spin-orbit coupled system is studied as an example,
and it is found that the intrinsic second order contribution plays a major role in the region of $\mu>0$, while current mainly comes from the extrinsic terms when $\mu<0$. 
% and the value of intrinsic spin conductivity exhibits a maximum at zero chemical potential $\mu$ and decreases rapidly with increasing $\mu$; while the extrinsic contribution remains unchanged in the region of $\mu \geqslant 0$.
Thus, the dependence of spin conductivity on chemical potential is expected to distinguish the extrinsic and intrinsic contributions experimentally.
%}
Our calculation provides a theoretical basis for understanding the nonlinear spin transport.
\end{abstract}

\maketitle

\section{INTRODUCTION}
The Hall effect plays an important role in condensed matter physics, which can induce a transverse charge current in a perpendicular magnetic field $\boldsymbol{B}$ \cite{Klitzing,Prange,Xiao}.
Owing to Onsager's reciprocal relations, the linear Hall conductivity is forbidden when the Hamiltonian is invariant under time-reversal symmetry (TRS) \cite{Onsager1931,Onsager1931a}.
Recently, three nonlinear charge responses beyond linear response theory have been proposed \cite{Wang2021,Zhang2023}.
The first one, proposed by Sodemann and Fu in 2015 \cite{Sodemann2015}, revealed that the Berry curvature  dipole (BCD) in the $\boldsymbol{k}$ space results in a nonlinear Hall response and can be realized in a class of materials with TRS but broken inversion symmetry (IS), such as few-layer WTe$_2$ \cite{Du2018,Qiong,Kang,Li2023,Ye2023,Wang2019}, strained graphene \cite{Battilomo2019,Pantaleon2021,Zhang2022} and so on.
Subsequently, the second one contributed by the Berry connection polarizability (BCP) was proposed in 2021 \cite{Wang2021,Liu2021}, which is also a Hall effect but independent of the relaxation time of electrons $\tau$, suggesting that the Joule heat  is zero. It should be noted that this class of nonlinear response needs to simultaneously break $\mathcal{P} $ and $\mathcal{T} $ symmetry and can be applicable in some antiferromagnets \cite{Wang2023,Gao2023,Wang2023a}.
Herein, the BCP represents the correction of the Berry connection by the external electric field, which is equal to the quantum metric divided by the energy difference in a two-band model \cite{Gao2014}.  Thus, in other words, this kind of contribution in Refs. \cite{Gao2023,Wang2023a} is attributed  to the quantum metric resulting in nonlinear response.
Furthermore, it is worth mentioning that there are other intrinsic contributions to the second-order current related to the quantum metric according to some different methods \cite{Nandy2019,Das2023,Oiwa2022,Kaplan2022,Michishita2022,Wang2024}, which are not discussed in this work.
According to Ref. \cite{Zhang2021}, the third class of contributions that can be traced back to the group velocity of electrons, known as Drude effect, exists both transversely and longitudinally.

%\textcolor{red}{
Apart from charge freedom, electrons possess spin degree which is crucial for spintronics. It is thus a basic question to look for mechanisms for topological nonlinear spin transport.
As a spin version of the quantum Hall effect, quantum spin Hall effect has been explored as a consequence of linear response \cite{Shen2017}.
%
%Similarly, it is of great significance to find the spin version of nonlinear charge current for spintronics, where maybe there should also be three different classes of nonlinear spin response.
%
%
Remarkably, a nonlinear spin current was proposed by Hamamoto \textit{et al.} \cite{Hamamoto2017}  in 2017, only taking into account the effect of group velocity on spin transport, which  is similar to the nonlinear charge Drude effect.
Subsequently, another spin Hall effect contributed by the spin-dependent BCD has been discussed in $\mathcal{PT} $ symmetric antiferromagnets metals has been discussed in 2022 \cite{Hayami2022}, which is induced by an effective spin-dependent hopping in AFM orderings without relying on spin-orbit coupling (SOC).
However, these two mechanisms are all extrinsic and depend on relaxation time, and the intrinsic mechanism has not been appreciated and exposed. 
On the other hand, none of the above  explains the effect of SOC on nonlinear spin Hall current.
Motivated by the above discussion, we are focusing on  identifying  a second-order intrinsic spin response mechanism besides the BCD and Drude-like mechanisms.
This would have a great impact on our understanding of the nonlinear response of spin.

In this work, starting from the conventional definition of spin current \cite{Murakami2004,Murakami2004a,Sinova2004}, we obtain a second-order intrinsic spin current contributed by the \textit{spin Berry curvature polarizability (SBCP)}, where the influence of the external electric field on the electron Bloch state is calculated using the standard perturbation theory. In addition, Drude and BCD-type terms of second-order spin currents are also discovered like its charge counterparts, from group velocity and spin-dependent Berry curvature dipole (SBCD).

To give a deeper and more specific understanding, we calculate the proposed second-order spin current in a two-dimensional model with Rashba-Dresselhaus SOC.
Intriguingly, the magnitude of the extrinsic contribution increases with the increase of the chemical potential, which is the main source of second order spin current only when $\mu<0$. However, the value of the intrinsic conductivity increases rapidly from zero to a finite value and remains unchanged when the chemical potential passes zero, which makes the SBCP is the dominant contribution in the region of $\mu>0$.

This indicates that we can distinguish the intrinsic and extrinsic spin currents by adjusting the level of the Fermi surface, which provides a certain information for experiments.

The structure of this paper is organized as follows. In Sec. \ref{sec2},
we develop a formalism and obtain three different kinds of second-order spin currents, and propose the second-order intrinsic spin Hall effect.
In Sec. \ref{sec3}, applying our theory to the two-dimension Rashba-Dresselhaus model, we discuss in detail the second-order spin currents and calculate their variations with chemical potential and model parameters. Finally, we summarize and discuss our results in Sec. \ref{sec4}.

\section{SECOND-ORDER intrinsic  SPIN  CURRENT}
\label{sec2}
\subsection{The definition of spin current}
The conventional spin current can be defined by the integration of the spin current operator and distribution function  \cite{Murakami2004,Murakami2004a,Sinova2004} as,
\begin{equation}
   J_{i}^{\ell}=\sum_n{\int_{\boldsymbol{k}} j_{i,n}^{\ell } }f_n,
    \label{eq1}
\end{equation}
where  $n$ is the band index, $\int_{\boldsymbol{k}} = \int{\frac{d^D \boldsymbol{k}}{(2\pi )^D}}$ with $D$ the dimension, and
$j_{i,n}^{\ell }=\langle n|\hat{j}_{i}^{\ell }|n\rangle (i=x,y;\ell=x,y,z)$ 
in which $i$ is the direction of spin current, $\ell$ shows the direction of the spin polarization. $\hat{j}_{i}^{\ell}=\frac{1}{4}\left\{ \frac{\partial \hat{H}}{\partial k_i},\sigma ^\ell \right\} $ is the spin current operator given by the anti-commutator of the group velocity and spin operator.
$f_n$ is  the Fermi-Dirac distribution for the $n$-th energy band.

Applying an external electric field $\boldsymbol{E}=\sum_j E_j \boldsymbol{e}_{j}, (j=x,y)$ to the system, the spin current can be expressed as
\begin{equation}
	J_i^\ell =\sigma_i^\ell  +\chi_{ij}^\ell  E_j+\Gamma_{ijl}^\ell  E_jE_l+ \cdots,
\end{equation}
where the $\sigma_{i}^\ell , \chi_{ij}^\ell $ and $ \Gamma_{ijl}^\ell $ are the zeroth, first and second-order conductivities for spin current, respectively. In order to calculate the conductivity, we assume that external electric field is weak so that the perturbation theory can be applied.

\subsection{The time-independent perturbation theory in external electric field}
Perturbed by an electric field $\boldsymbol{E}$ with $H'=e\boldsymbol{E}  \cdot (\boldsymbol{r}-\boldsymbol{r}_c) $, the single-particle wave function in momentum space is written as
\begin{equation}
|\tilde{n}\rangle =|n\rangle +\sum_{m\ne n}{\frac{e \boldsymbol{E}\cdot \boldsymbol{A}_{mn}}{\varepsilon _{n}^{(0)}-\varepsilon _{m}^{(0)}}}|m\rangle ,
\label{eq3}
\end{equation}
where $\boldsymbol{A}_{mn} =i \langle m |  \nabla _{\boldsymbol{k}} |n \rangle  $ is the interband Berry connection, and $e$ is the charge of electron.
Then, the first-order modified Berry connection is obtained as,
\begin{equation}
	\tilde{A}_{i,nn}=A_{i,nn}+eG_{ij,n}E_j,
	\label{eq4}
\end{equation}
where $ G_{ij,n}=2\mathrm{Re}%\left[
\sum_{m\ne n}{\frac{A_{i,nm}{A}_{j,mn}}{\varepsilon _{n}^{(0)}-\varepsilon _{m}^{(0)}}} %\right]
$
is the Berry connection polarizability (BCP) \cite{Gao2014,Gao2019}.
Next, we use Eq. (\ref{eq3}) to obtain the modified spin current operator  reserved to the second-order of the electric field
\begin{equation}
\tilde{j}_{i,n}^{\ell}=\langle \tilde{n}|\hat{j}_{i,n}^{\ell}|\tilde{n}\rangle =j_{i,n}^{\ell}-e \tilde{\Omega}_{ij,n}^{\ell}E_j,
\label{eq5}
\end{equation}
where
\begin{equation}
\tilde{\Omega}_{ij,n}^{\ell} =\Omega _{ij,n}^{\ell}-e \Pi_{ijl,n}^{\ell}E_l.
\label{eq6}
\end{equation}
Here, $\Omega _{ij,n}^{\ell}=2 \mathrm{Im}
\sum_{m\ne n}{\frac{\langle n|\hat{j}_{i}^{\ell}|m\rangle \langle m|\frac{\partial H}{\partial k_j}|n\rangle}{\left( \varepsilon _{n}^{(0)}-\varepsilon _{m}^{(0)} \right) ^2}}$ is the spin-dependent Berry curvature (SBC), which can also be obtained by linear response theory \cite{Sinova2004,Yao2004,Guo2005,Guo2008,Chen2023}. Further, we define $\Pi_{ijl,n}^{\ell}$ as spin Berry curvature polarizability (SBCP),
\begin{equation}
	\begin{aligned}
\Pi _{ijl}^{\ell}=\sum_{\substack{m\ne n\\q\ne n}}{\frac{\langle u_{q}^{(0)}|\hat{j}_{i}^{\ell}|u_{m}^{(0)}\rangle \langle u_{n}^{(0)}|\frac{\partial H}{\partial k_j}|u_{q}^{(0)}\rangle \langle u_{m}^{(0)}|\frac{\partial H}{\partial k_l}|u_{n}^{(0)}\rangle}{\left( \varepsilon _{n}^{(0)}-\varepsilon _{q}^{(0)} \right) ^2\left( \varepsilon _{n}^{(0)}-\varepsilon _{m}^{(0)} \right) ^2}}.
		\label{eq7}
	\end{aligned}
\end{equation}
%\textcolor{red}{
As will be made clear below, this term leads to an intrinsic second-order spin current, which is one of main results in this work.
%Clearly, our main idea is that this term will contribute an intrinsic second-order spin current, which we will discuss later.
%}

Next, we drive the distribution function with the Boltzmann equation by using relaxation time approximation \cite{Mahan,Zhang2021},
\begin{equation}
	-\frac{e \boldsymbol{E}}{\hbar}\cdot \frac{\partial f}{\partial \boldsymbol{k}}=-\frac{f-f^{(0)}}{\tau},
\end{equation}
in which $\tau$ is the relaxation time for electrons, $f^{(0)}$ is the equilibrium distribution function without $\boldsymbol{E}$. The distribution function can be obtained by expanding the order of electric field ,
\begin{equation}
	\begin{aligned}
			f &=\sum_p f^{(p)} =\sum_p \left(\frac{e \tau}{\hbar}\right)^p\frac{\partial^p f^{(0)}}{\partial k_i^p} E_i^p (p=1,2,3,\cdots),\\
		%F^{(m)}&=\left(\frac{e \tau}{\hbar}\right)^m\frac{\partial^m f^{(0)}}{\partial k_i^m},\\
	\end{aligned}
\label{eq9}
\end{equation}
where $p$ in $f^{(p)}$ and $E_i^p$ means the order of distribution function and the power of $E_i$ respectively, and $i$ is determined by the direction of electric field.

\subsection{Linear and nonlinear spin current}
Motivated by the above discussions, we get the general spin current
\begin{equation}
	J_{i}^{\ell}=\sum_n  \sum_p \int_{\boldsymbol{k}} \left[ \langle u_{n}^{(0)}|\hat{j}_{i}^{\ell}|u_{n}^{(0)}\rangle -e\tilde{\Omega}_{ij,n}^{\ell}E_j \right] f^{(p)}.
\end{equation}
Each order of the currents is assembled by the sum of the superscript $p$. We omit the band index and reserve to second-order spin current.  The spin  conductivities are (see Appendix \ref{AppendixA})
%\begin{equation}
	\begin{align} \label{eq11}
	\sigma _{i}^{\ell} &=\int_{\boldsymbol{k}}{j_{i}^{\ell}}f^{(0)}, \notag
\\
\chi _{ij}^{\ell}&=e\int_{\boldsymbol{k}}{\left[ j_{i}^{\ell}\frac{\tau}{\hbar}\frac{\partial f^{(0)}}{\partial k_j}-\Omega _{ij}^{\ell}f^{(0)} \right]},
\\
\Gamma _{ijl}^{\ell}&=e^2\int_{\boldsymbol{k}}{\left[ j_{i}^{\ell}\frac{\tau ^2}{\hbar ^2}\frac{\partial ^2f^{(0)}}{\partial k_j\partial k_l}-\frac{\tau}{\hbar}\Omega _{ij}^{\ell}\frac{\partial f^{(0)}}{\partial k_l}+\Pi _{ijl}^{\ell}f^{(0)} \right]}. \notag
	\end{align}

The first line $ \sigma_i^\ell  $ indicates spin current does not vanish even in thermodynamic equilibrium in the absence of external fields \cite{Rashba2003}. For calculating transport spin currents, a procedure of eliminating the background currents should be devised.
The second line, $\chi_{ij}^\ell $, shows linear spin conductivity for spin current, of which the second term describes the intrinsic linear spin Hall effect contributed by the SBC in the Kubo formula \cite{Guo2005,Feng2012}.
Importantly, the last term of $\Gamma _{ijl}^{\ell}$ is critical in this work, which contributes to the intrinsic second-order spin current induced by the SBCP, independent of relaxation time $\tau$, and can be rewritten as $\Gamma_{ijl}^{\ell, \text{int}}$. 
In passing, the first term is coming from group velocity, which has be proposed in Ref. \cite{Hamamoto2017}. After integration by parts, the second term can be written as the integral of the product of partial of SBC and distribution function, being defined as SBCD. Meanwhile, we note that these two terms are both related to $\tau$, and can be called as extrinsic contributions $\Gamma_{ijl}^{\ell, \text{ext}}$.

\section{THE MODEL WITH SOC}
\label{sec3}
The Hamiltonian including the Rashba-type and Dresselhaus-type SOC is
\begin{equation}
	H_{\text{RD}}=\frac{\hbar^2 k^2}{2m}+\alpha (k_x\sigma^y-k_y\sigma^x)+\beta (k_x\sigma ^x-k_y\sigma^y),
	\label{eq12}
\end{equation}
where $\alpha$ and $\beta$ indicate the strength of interactions, $\sigma^{x,y}$ are Pauli matrices. The energy dispersion is  $\varepsilon _{\pm}=\frac{ \hbar^2  k^2}{2m}\pm kA(\phi)$,   where  $k=\sqrt{k_x^2+k_y^2}$,\,$k_x=k\cos\phi,\,k_y=k\sin\phi$ and $A(\phi)= \sqrt{\left( \alpha ^2+\beta ^2 \right) -2\alpha \beta \sin2\phi}$.
The spin polarization in the $k$ space is $\langle \pm |\boldsymbol{\sigma }|\pm \rangle =\pm \left( \cos \varphi ,-\sin \varphi ,0 \right) $, where $\varphi =\text{arg}\left[ \left( \beta k_x-\alpha k_y \right) +i\left( \beta k_y-\alpha k_x \right) \right] $.
The chemical potential is set to be $\mu$, and the wave vectors on the Fermi surface are  $k_{F-}^+$ and $k_{F+}^+$ when $\mu>0$ while $k_{F-}^+$ and $k_{F-}^-$ when $\mu<0$.
In order to get closer to the real materials, we impose a condition $\alpha^2+\beta^2 =1$, i.e., $\alpha=|\cos\theta|$ and $\beta=\sin \theta$ $(\theta =[0,\pi/2] )$ \cite{Hamamoto2017}.

%%%%%%%%%%%%%%%%%%%%%%%%%%%%%%%%%%
 \begin{figure}[th]
	\includegraphics[width=1\linewidth]{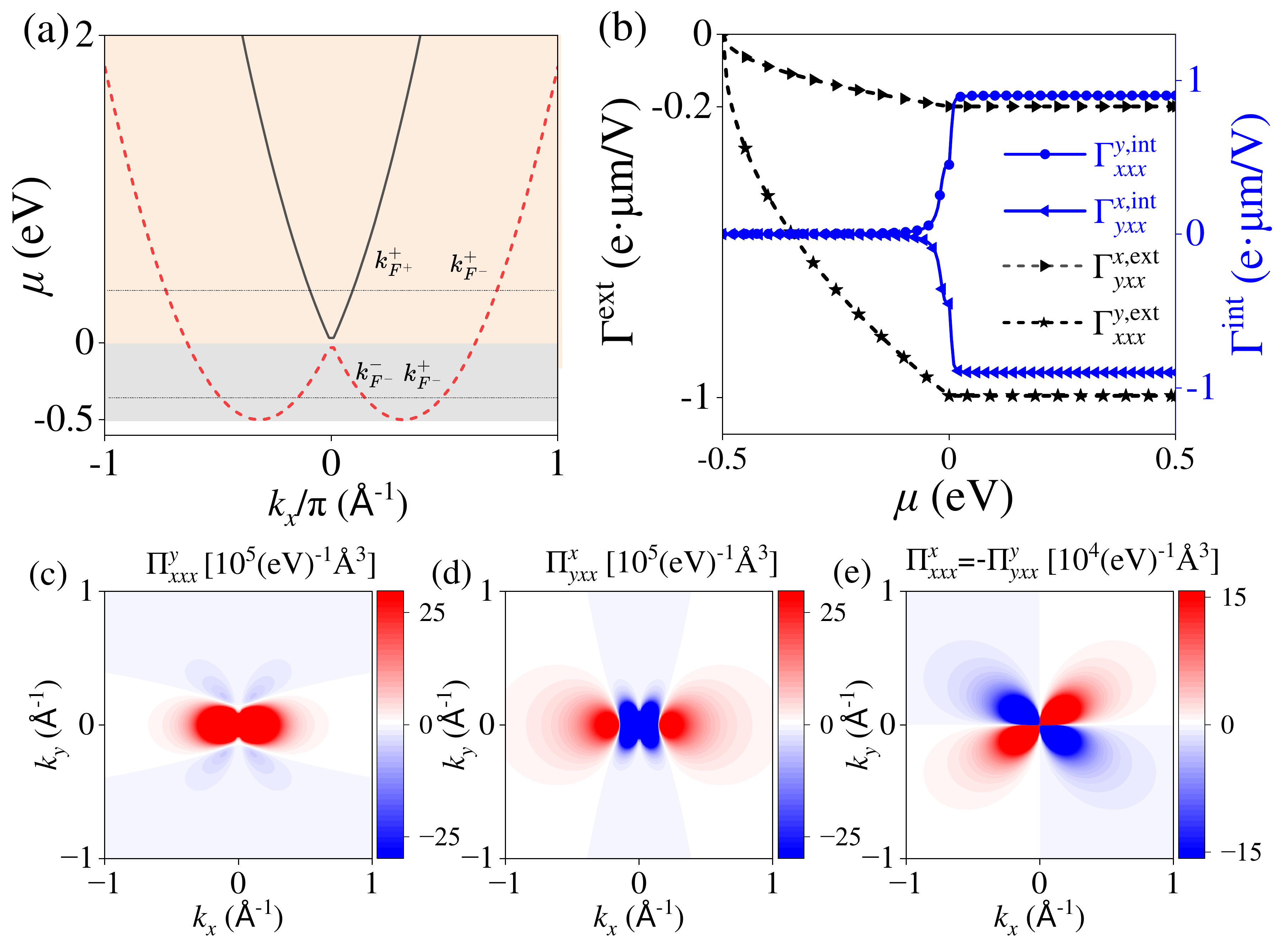}
	\caption{For the pure Rashba system with $\theta=0$ ($\alpha=1$ and $\beta=0$). (a) Two-band structure where the $k_{F^{\pm}}^\pm$ are wave vectors in the Fermi surface, and the light orange and light gray indicate $\mu>0$ and $\mu<0$, respectively. (b) The second-order spin conductivity $\Gamma _{yxx}^{x,\text{ext/int}}$ and $\Gamma _{xxx}^{y, \text{ext/int}}$ vs chemical potential $\mu$. (c)-(e) 2D contour plot of the SBCP $\Pi _{ixx,+}^{j} $.  }
	\label{fig1}
\end{figure}
%%%%%%%%%%%%%%%%%%%%%%%%%%%%%%%%%%%%%%%%%%%%%%%%%%%%%%%%%%%%%%%%%%%%%%%%%%%%%%%%%%%%%%%%%%%%%%%%%%%%%%%%%%%%%%%
According to Eq. (\ref{eq11}), there should also be non-transport spin current  $\sigma_i^\ell$ in this model \cite{Rashba2003}, which we will not discuss here (see Appendix \ref{appendixB}).
Meanwhile, spin-dependent Berry curvature, like Berry curvature, is zero everywhere for  gapless systems, expect for band divergence at the point of $k_x =k_y =0$. Thus, there is no spin current induced by spin Berry curvature whatever first order or second order. Importantly, this model (Eq. (\ref{eq12})) satisfies the TRS but breaks IS, resulting in all the linear spin currents vanish after the angle $\phi$ integral. Thus, it is very appropriate to account for intrinsic and extrinsic second-order contributions. On the other hand, we only consider the case of second-order spin currents generated by in-plane polarization due to  $\langle \pm |\sigma^z |\pm \rangle =0 $.

For simplicity, we assume that the applied electric field is along the $x$ direction, and we get the SBCP (see Appendix \ref{appendixB} for more details)
\begin{equation}
    \Pi _{ixx,\pm}^{\ell}=\frac{\left( \alpha ^2-\beta ^2 \right) ^2k_{y}^{2}}{32 mk^7A^7(\phi )}\left(\mp \hbar^2 \hat{\mathcal{K}}  +mkA(\phi)\hat{I} \right) \hat{\mathcal{D}}, \\
\end{equation}
where $\hat{I}$ is a $2\times2$ unit matrix, $\hat{\mathcal{K}}=\left[ \begin{matrix}
	k_{x}^{2}&		k_xk_y\\
	k_xk_y&		k_{y}^{2}\\
\end{matrix} \right]$ and $\hat{\mathcal{D}}=\left[\begin{array}{cc}\beta,\alpha\\ -\alpha,-\beta\end{array}\right] $ being just a matrix of coefficients related to the SOC parameters.

%\subsection{The Rashba-Dresselhuas system}

%\subsection{Pure Rashba system}
For the case of only Rashba SOC, the Hamiltonian reduces to a pure Rashba system for $\theta=0$ ($\alpha=1$ and $\beta=0$). As shown in Fig. \ref{fig1}(a), the two bands are degenerate at the $k_x =k_y =0 $, and we have marked the approximate positions of $ k_{F^\pm}^\pm$. The second-order spin conductivities are derived with four terms, $\Gamma_{xxx}^{y,\text{ext/int}}$ and $\Gamma_{yxx}^{x,\text{ext/int}}$, shown in Fig. \ref{fig1}(b).

For extrinsic contributions, $\Gamma_{xxx}^{y, \text{ext}}= 5 \Gamma_{yxx}^{x,\text{ext}}  $, the magnitude of spin conductivity $\Gamma_{xxx}^{y, \text{ext}}$ increases from zero at $\mu = -0.5$ to a saturation $-0.1$ around $\mu$ = 0 and almost keeps this value unchanged for $\mu>0$.
For better analysis of intrinsic contributions, we depict a 2D project of the SBCP on the $(k_x,k_y)$ plane, where $\Pi_{xxx}^{x}$ and $\Pi_{yxx}^{y}$  are odd, $\Pi_{yxx}^{x}$ and $\Pi_{xxx}^{y}$ are even [Fig. \ref{fig1}(c)-(e)]. Thus, we can make it clear that there are also two intrinsic contributions, and the direction spin and current are always perpendicular due to spin-momentum locking in pure Rashba systems. From the quantitative relationship,  we can obtain that when $\mu<0$ (away from zero), the extrinsic term $\Gamma_{yxx}^{x, \text{ext}}$ gradually decreases in magnitude with the increase of $|\mu|$ while when $\mu>0$ it almost does not vary with $\mu$.
As for the intrinsic term, we have $\Gamma_{yxx}^{x, \text{int}}=-\Gamma_{xxx}^{y, \text{int}}$, whose magnitudes begin to increase rapidly near  the point of band degeneracy $\mu =0$, and then continue to a fixed value as $\mu >0$. From the analysis of SBCP [Fig. 1 (c-e)], we find that $\Pi_{xxx,+}^y$ and $\Pi_{yxx,+}^x$ diverge at the point of $\mu=0$, which  causes the current increases rapidly  when the Fermi level passes through the degeneracy point. 
Similarly,  both intrinsic and extrinsic spin current tend to zero at $\mu = -0.5$, which may be understood as that the nontrivial Berry curvature texture almost around the band maximum and minimum at $k_{x}=k_{y}=0$ and does not enter into the transport when $\mu$ lies deeply in the lower band.
On the other hand, the intrinsic contributions influence both transverse and longitudinal spin current in the region of $\mu>0$. For the case of spin current along the $x$ direction, $\Gamma_{xxx}^{y,\text{int}}$ and $\Gamma_{xxx}^{y,\text{ext}}$ are the opposite but of similar value, which means that the intrinsic term will compensate for the longitudinal spin current to almost zero. For the transverse  term, intrinsic term effectively increases the total spin current, which mainly come from SBCP. Thus, it is hopeful that the characteristics of second-order intrinsic spin current can be investigated by measurement when the appropriate chemical potential is adjusted.
%%%%%%%%%%%%%%%%%%%%%%%%%%%%%%%%%%%%%%%%%%%%%%%%%%%%%%%%%%%%%%%%%%%%
\begin{figure}[tb]
	\includegraphics[width=1\linewidth]{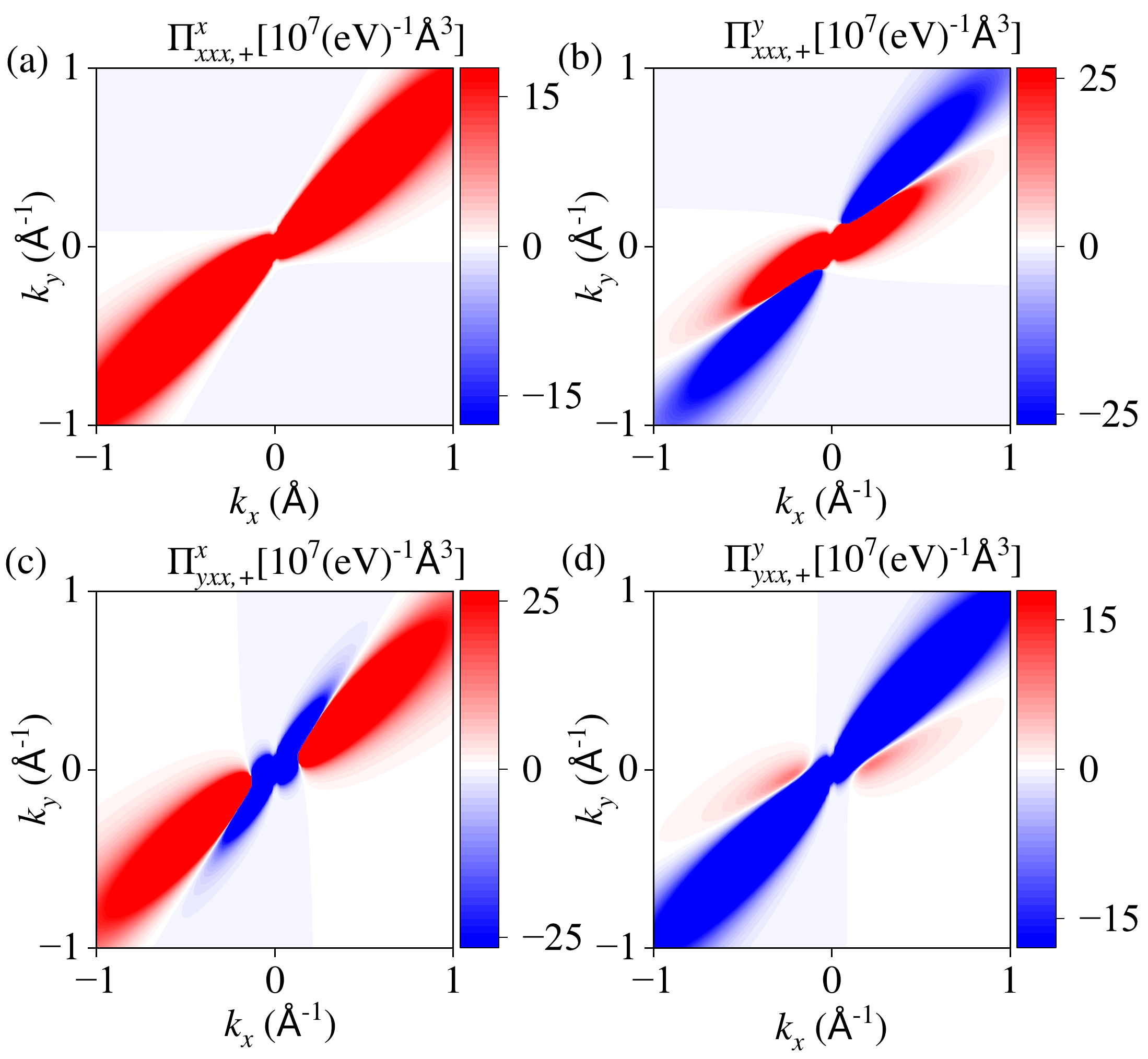}
	\caption{The 2D contour plot for SBCP in the case of $\theta =\pi/6$ ($\alpha = \sqrt{3}/2, \beta=1/2$) \textcolor{red}{only for $+$ band} where (a) $\Pi_{xxx,+}^{x}$, (b) $\Pi_{xxx,+}^{y}$, (c) $\Pi_{yxx,+}^{x}$, (d) $\Pi_{yxx,+}^{y}$. }
	\label{fig2}
\end{figure}
%%%%%%%%%%%%%%%%%%%%%%%%%%%%%%%%%%%%%%%%%%%%%%%%%%%%%%%%%%%%%%%%%%%%

It is important to note that the other four components of the spin conductivity exist in pure Dresselhaus system with $\theta=\pi/2$ ($\alpha =0$ and $\beta =1$), $\Gamma_{xxx}^{x, \text{ext/int}}$ and $ \Gamma_{yxx}^{y, \text{ext/int}}$.
We give a detailed derivation and description in the Appendix \ref{appendixB}.

%%%%%%%%%%%%%%%%%%%%%%%%%%%%%%%%%%%%%%%%%%%%%%%%%%%%%%%%%%%%%%%%%%%%

\begin{figure}[tb]
	\includegraphics[width=1\linewidth]{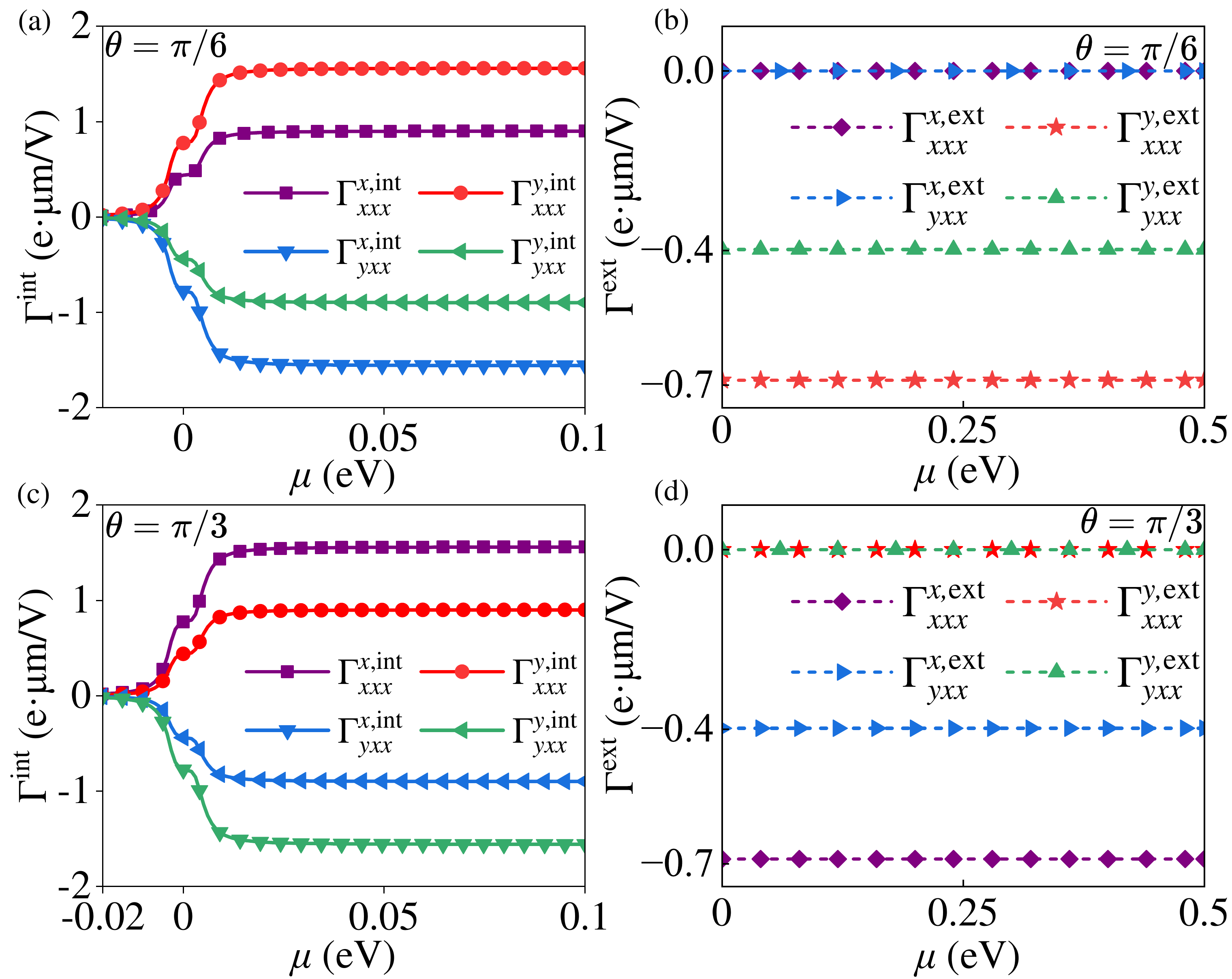}
	\caption{The intrinsic and extrinsic spin conductivities vs chemical potential $\mu$ in the case of (a-b) $\theta =\pi/6$ ($\alpha = \sqrt{3}/2, \beta=1/2$) and (c-d) $\theta =\pi/3$ ($\alpha = 1/2, \beta=\sqrt{3}/2$) v, where (a) and (c) for $\Gamma^{\text{int}}$, (b) and (d) for  $\Gamma^{\text{ext}}$.}
	\label{fig3}
\end{figure}

%%%%%%%%%%%%%%%%%%%%%%%%%%%%%%%%%%%%%%%%%%%%%%%%%%%%%%%%%%%%%%%%%%%%
According to our analysis, in the present of both Rashba and Dressselhaus SOC, all four components of the intrinsic spin conductivities are non-zero except in the degenerate case with $\alpha = \beta$. To illustrate this effect, we choose two different cases, $\theta=\pi/6$ and $\pi/3$, to discuss. Figure \ref{fig2} describer the distribution of SBCP in $k_x-k_y$ plane when $\theta=\pi/6$ ($\alpha=\sqrt{3}/2, \beta=1/2$), implying that they are both maximized or minimized at the origin point $k_x = k_y =0$.
In Fig. \ref{fig3}, we show the dependence of the intrinsic and extrinsic spin conductivities on the chemical potential $\mu$.
Take the case of $\theta =\pi/6$ as an example, the magnitudes of all non-vanishing spin currents increase and tend to their stable values respectively with increasing $\mu$ in positive region, which is similar to the intrinsic linear Hall effect in a gaped model \cite{Sinova2004}.
It is shown in Fig. \ref{fig3}(a) that all four intrinsic spin conductivities exist, but $\Gamma_{xxx}^{x/y, \text{int}} $ are positive and  $\Gamma_{yxx}^{x/y, \text{int}} $ are negative.
From Fig. \ref{fig3} (b), there are only two non-vanishing second-order extrinsic spin currents, i.e. $\Gamma _{yxx}^{y, \text{ext}} = (4/7) \Gamma _{xxx}^{y, \text{ext}}$,  which are constants when tuning $\mu$. 
When $\theta =\pi/3$, $\alpha$ and $\beta$ exchange, and a similar situation occurs for non-zeros second-order spin conductivity [Fig. \ref{fig3} (c)-(d)].

\section{Conclusion and Discussion}
\label{sec4}
In this work, we propose a mechanism to second-order intrinsic spin current contributed by the SBCP, independent of the relaxation time.
Furthermore, we complete other two kinds of contributions of second-order spin response, which stem from group velocity and SBCD, respectively.
To illustrate the effect of SOC, we apply our theory to a two-band Rashba-Dresselhaus model.
In the region of $\mu>0$, the spin  current is mainly caused by SBCP while the extrinsic contribution dominates when $\mu<0$.
This will provide us with an effective way to distinguish extrinsic and intrinsic contributions. %Meanwhile, we also 

%%%%%%%%%%%%%%%%%%%%%%%%%%%%%%%%%%%%%%%%%%%%%%%%%%%%%%%%%%%%%%%%%%%%
%%%%%%%%%%%%%%%%%%%%%%%%%%%%%%%%%%%%%%%%%%%%%%%%%%%%%%%%%%%%%%%%%%%%

In addition, there are a few different approaches to discuss second order spin current.
Ref. \cite{Kapri2021} study the background (equilibrium or zeroth-order), linear (first-order) and nonlinear (second-order) spin currents response in two-dimensional Rashba
spin-orbit coupled systems with Zeeman splitting to emphasize the role of Berry curvature in spin transport.
They use a modified spin current operator by inclusion of the anomalous velocity, which is similar to the contribution of  SBC in our work. Importantly, here we mainly discuss the effect of SBCP  to obtain an intrinsic second-order spin current.
Another work \cite{MunozSantana2023} proposed the linear and the second harmonic spin current response in two anisotropic systems with spin-orbit interaction using the Kubo response theory. 
And it is  classified in Ref. \cite{Lihm2022} that the second-order spin photocurrents are divided into Drude, Berry curvature
dipole, shift, injection, and rectification currents.In our work, only the dc second-order intrinsic spin current is discussed using the correction of the Bloch wave function by the external electric field.
It should be noted that the second order spin current we proposed contains both transverse and longitudinal contributions, which is interesting to be a next study. 
Moreover, if the ac response is taken into account, it is in principle possible to obtain second-order spin current of finite frequency. 
What's more, there is no further symmetry analysis for the second order spin current, which allows us to quickly search for possible materials and we hope to improve it in the future work.

\begin{acknowledgments}
The authors thank Dr. Yuan-dong Wang for fruitful discussions. This work is supported in part by the Training Program of Major Research plan of the National Natural Science Foundation of China (Grant No. 92165105), and CAS Project for Young Scientists in Basic Research Grant No. YSBR-057.  It is supported by the Strategic Priority Research Program of CAS (Grant No. XDB28000000, and No. XDB33000000). And it is also supported in part by the NSFC (Grant No. 11974348 and No. 11834014).
\end{acknowledgments}

\appendix
\setcounter{figure}{0}
\renewcommand{\thefigure}{A\arabic{figure}}
\renewcommand{\thetable}{A\arabic{table}}
\begin{widetext}
\section{The second-order intrinsic spin current}
\label{AppendixA}
\subsection{Derivation of SBCP}
Perturbed by an electric field $\boldsymbol{E}$ with $H'=e\boldsymbol{E}  \cdot (\boldsymbol{r}- \boldsymbol{r_C})$,
Then, the first-order modified energy is obtained as,
\begin{equation}
    \varepsilon_{n}^{(1)}=\langle u_{n}^{(0)}|H^{\prime}|u_{n}^{(0)}\rangle =\langle u_{n}^{(0)}|eE\cdot \left( r-r_C \right) |u_{n}^{(0)}\rangle =0,
\end{equation}
and  the single-particle wave function in momentum space is written as
\begin{equation}
	\tilde{\boldsymbol{A}}_{mn} = \langle \tilde{n}|i\nabla _{\boldsymbol{k}}|\tilde{n}\rangle
	=  \langle n|i\nabla _{\boldsymbol{k}}|n\rangle +\sum_{m\ne n}{\left( \frac{e \boldsymbol{E}\cdot \boldsymbol{A}_{mn}}{\varepsilon _{n}^{(0)}-\varepsilon _{m}^{(0)}} \right) ^*}\langle m|i\nabla _{\boldsymbol{k}}|n\rangle
	+\sum_{m\ne n}{\frac{e \boldsymbol{E}\cdot \boldsymbol{A}_{mn}}{\varepsilon _{n}^{(0)}-\varepsilon _{m}^{(0)}}}\langle n|i\nabla _{\boldsymbol{k}}|m\rangle .
\end{equation}
Next we use the relation about the Berry connection $	\boldsymbol{A}_{mn} =\boldsymbol{A}_{nm}^{*}$. Then,
\begin{equation}
	\tilde{A}_{i,mn}=A_{i,mn}+eG_{ij}E_j,
\end{equation}
where $ 	G_{ij} = 2 \mathrm{Re} \left[\sum _{m\neq n}\frac{A_{i,nm} A_{j,mn}}{\varepsilon _{n}^{( 0)} -\varepsilon _{m}^{( 0)}}\right]$  is the Berry connection polarizability (BCP).

The second-order modified energy is
\begin{equation}
    \varepsilon _{n}^{(2)}=\sum_{m\ne n}{\frac{\langle u_{m}^{(0)}|H^{\prime}|u_{n}^{(0)}\rangle \langle u_{n}^{(0)}|H^{\prime}|u_{m}^{(0)}\rangle}{\varepsilon _{n}^{(0)}-\varepsilon _{m}^{(0)}}} =\sum_{m\ne n}{\frac{e^2\left( E\cdot A_{mn} \right) \left( E\cdot A_{nm} \right)}{\varepsilon _{n}^{(0)}-\varepsilon _{m}^{(0)}}} =\frac{1}{2}e^2G_{jl}E_jE_l.
\end{equation}
The spin current operator is
\begin{equation}\label{eq}
\begin{aligned}
\tilde{j}_{i,n}^{\ell} = &\langle u_n|\hat{j}_{i}^{\ell}|u_n\rangle =\left[ \langle u_{n}^{(0)}|+\sum_{q\ne n}{\frac{e\boldsymbol{E}\cdot \boldsymbol{A}_{qn}^{*}}{\varepsilon _{n}^{(0)}-\varepsilon _{q}^{(0)}}}\langle u_{q}^{(0)}| \right] \hat{j}_{i}^{\ell}\left[ |u_{n}^{(0)}\rangle +\sum_{m\ne n}{\frac{e\boldsymbol{E}\cdot \boldsymbol{A}_{mn}}{\varepsilon _{n}^{(0)}-\varepsilon _{m}^{(0)}}}|u_{m}^{(0)}\rangle \right]
\\
= &\langle u_{n}^{(0)}|\hat{j}_{i}^{\ell}|u_{n}^{(0)}\rangle +\sum_{q\ne n}{\frac{e\boldsymbol{E}\cdot \boldsymbol{A}_{nq}^{}}{\varepsilon _{n}^{(0)}-\varepsilon _{q}^{(0)}}}\langle u_{q}^{(0)}|\hat{j}_{i}^{\ell}|u_{n}^{(0)}\rangle +\sum_{m\ne n}{\frac{e\boldsymbol{E}\cdot \boldsymbol{A}_{mn}}{\varepsilon _{n}^{(0)}-\varepsilon _{m}^{(0)}}}\langle u_{n}^{(0)}|\hat{j}_{i}^{\ell}|u_{m}^{(0)}\rangle \\
&+\sum_{q\ne n}{\frac{e\boldsymbol{E}\cdot \boldsymbol{A}_{nq}^{}}{\varepsilon _{n}^{(0)}-\varepsilon _{q}^{(0)}}}\sum_{m\ne n}{\frac{e\boldsymbol{E}\cdot \boldsymbol{A}_{mn}}{\varepsilon _{n}^{(0)}-\varepsilon _{m}^{(0)}}}\langle u_{q}^{(0)}|\hat{j}_{i}^{\ell}|u_{m}^{(0)}\rangle ,
\end{aligned}
\end{equation}
where the second and third term in the second line can be merged,
\begin{equation}\label{eq}
\begin{aligned}
 &\sum_{q\ne n}{\frac{e\boldsymbol{E}\cdot \boldsymbol{A}_{nq}^{}}{\varepsilon _{n}^{(0)}-\varepsilon _{q}^{(0)}}}\langle u_{q}^{(0)}|\hat{j}_{i}^{\ell}|u_{n}^{(0)}\rangle +\sum_{m\ne n}{\frac{e\boldsymbol{E}\cdot \boldsymbol{A}_{mn}}{\varepsilon _{n}^{(0)}-\varepsilon _{m}^{(0)}}}\langle u_{n}^{(0)}|\hat{j}_{i}^{\ell}|u_{m}^{(0)}\rangle \\
& =e\sum_{m\ne n}{\frac{\boldsymbol{E}\cdot \boldsymbol{A}_{nm}\langle u_{m}^{(0)}|\hat{j}_{i}^{\ell}|u_{n}^{(0)}\rangle +\boldsymbol{E}\cdot \boldsymbol{A}_{mn}\langle u_{n}^{(0)}|\hat{j}_{i}^{\ell}|u_{m}^{(0)}\rangle}{\varepsilon _{n}^{(0)}-\varepsilon _{m}^{(0)}}}
\\
&=e\sum_{m\ne n}{2\mathrm{Re}\left[ \frac{\boldsymbol{A}_{mn}^{j}\langle u_{n}^{(0)}|\hat{j}_{i}^{\ell}|u_{m}^{(0)}\rangle}{\varepsilon _{n}^{(0)}-\varepsilon _{m}^{(0)}} \right]} E_j
=e\hbar \sum_{m\ne n}{2\mathrm{Re}\left[ \frac{i\langle u_{n}^{(0)}|\hat{j}_{i}^{\ell}|u_{m}^{(0)}\rangle \langle u_{m}^{(0)}|v_j|u_{n}^{(0)}\rangle}{(\varepsilon _{n}^{(0)}-\varepsilon _{m}^{(0)})^2} \right]}E_j
\\
&=-e\hbar \sum_{m\ne n}{2\mathrm{Im}\left[ \frac{\langle u_{n}^{(0)}|\hat{j}_{i}^{\ell}|u_{m}^{(0)}\rangle \langle u_{m}^{(0)}|v_j|u_{n}^{(0)}\rangle}{(\varepsilon _{n}^{(0)}-\varepsilon _{m}^{(0)})^2} \right]}E_j
=-e\Omega_{ij}^\ell E_j.
\end{aligned}
\end{equation}
And the
\begin{equation}
  \Omega _{ij}^{\ell}=\sum_{m\ne n}{2\mathrm{Im}\left[ \frac{\langle u_{n}^{(0)}|\frac{1}{2}\left\{ \frac{\partial H}{\partial k_i},\sigma ^{\ell} \right\} |u_{m}^{(0)}\rangle \langle u_{m}^{(0)}|\frac{\partial H}{\partial k_j}|u_{n}^{(0)}\rangle}{(\varepsilon _{n}^{(0)}-\varepsilon _{m}^{(0)})^2} \right]},
\label{eqs9}
\end{equation}
is the spin-dependent Berry curvature and  its dimension is $L^2$.
The fourth term is the second-order modified,
\begin{equation}\label{eq}
\begin{aligned}
 &\sum_{q\ne n}{\frac{e\boldsymbol{E}\cdot \boldsymbol{A}_{nq}^{}}{\varepsilon _{n}^{(0)}-\varepsilon _{q}^{(0)}}}\sum_{m\ne n}{\frac{e\boldsymbol{E}\cdot \boldsymbol{A}_{mn}}{\varepsilon _{n}^{(0)}-\varepsilon _{m}^{(0)}}}\langle u_{q}^{(0)}|\hat{j}_{i}^{\ell}|u_{m}^{(0)}\rangle =e^2\sum_{m\ne n,q\ne n}{\frac{\left( \boldsymbol{E}\cdot \boldsymbol{A}_{nq}^{} \right) \left( \boldsymbol{E}\cdot \boldsymbol{A}_{mn}^{} \right)}{\left( \varepsilon _{n}^{(0)}-\varepsilon _{q}^{(0)} \right) \left( \varepsilon _{n}^{(0)}-\varepsilon _{m}^{(0)} \right)}}\langle u_{q}^{(0)}|\hat{j}_{i}^{\ell}|u_{m}^{(0)}\rangle
\\
&=e^2  \sum_{m\ne n,q\ne n}{\frac{\left( \boldsymbol{E}\cdot \boldsymbol{A}_{nq}^{} \right) \left( \boldsymbol{E}\cdot \boldsymbol{A}_{mn}^{} \right)}{\left( \varepsilon _{n}^{(0)}-\varepsilon _{q}^{(0)} \right) \left( \varepsilon _{n}^{(0)}-\varepsilon _{m}^{(0)} \right)}}\langle u_{q}^{(0)}|\hat{j}_{i}^{\ell}|u_{m}^{(0)}\rangle \\
&=e^2\hbar^2 \sum_{m\ne n,q\ne n}{\frac{\langle u_{q}^{(0)}|\hat{j}_{i}^{\ell}|u_{m}^{(0)}\rangle \langle u_{n}^{(0)}|v_j|u_{q}^{(0)}\rangle \langle u_{m}^{(0)}|v_l|u_{n}^{(0)}\rangle}{\left( \varepsilon _{n}^{(0)}-\varepsilon _{q}^{(0)} \right) ^2\left( \varepsilon _{n}^{(0)}-\varepsilon _{m}^{(0)} \right) ^2}}E_jE_l
\\
& =e^2 \Pi_{ijl}^\ell E_j E_l.
\end{aligned}
\end{equation}
If we only consider a two-band model, we will have $q = m$,
\begin{equation}
   \Pi _{ijl}^{\ell}=\frac{\langle u_{m}^{(0)}|\hat{j}_{i}^{\ell}|u_{m}^{(0)}\rangle \langle u_{n}^{(0)}|\frac{\partial H}{\partial k_j}|u_{m}^{(0)}\rangle \langle u_{m}^{(0)}|\frac{\partial H}{\partial k_l}|u_{n}^{(0)}\rangle}{\left( \varepsilon _{n}^{(0)}-\varepsilon _{m}^{(0)} \right) ^4}.
\end{equation}

\subsection{Derivation of linear and nonlinear spin conductivity}
According to Eq. (\ref{eq1}), when we know the specific forms of the spin current operator $\hat{j}_{i}^{\ell }$ [Eq. (\ref{eq5})], and the distribution function [Eq. (\ref{eq9})], we can get the total spin current containing any order,
\begin{equation}\label{eq}
\begin{aligned}
J_{i}^{\ell} &=\sum_n{\sum_p{\int_{\boldsymbol{k}}{\tilde{j}_{i,n}^{\ell}}}}f^{(p)}=\sum_n{\sum_p{\int_{\boldsymbol{k}}{\left( j_{i,n}^{\ell}-e\tilde{\Omega}_{ij,n}^{\ell}E_j \right)}}}\left( \frac{e\tau}{\hbar} \right) ^p\frac{\partial f^{(0)}}{\partial k_{i}^{p}}E_{i}^{p}
\\
&=\sigma _{i}^{\ell}+\chi _{ij}^{\ell}E_j+\Gamma _{ijl}^{\ell}E_jE_l+\cdots.
\end{aligned}
\end{equation}
where $p$ is the order of spin current.  Next we will solve the current of each order where we ignore the energy band index. When $p=0$,
\begin{equation}
    J_{i}^{\ell \left( p=0 \right)}=\int_{\boldsymbol{k}}{j_{i}^{\ell}f^{(0)}}=\sigma _{i}^{\ell}.
\label{eqA11}
\end{equation}
When $p=1$,
\begin{equation}
    J_{i}^{\ell \left( p=1 \right)}=\int_{\boldsymbol{k}}{j_{i}^{\ell}f^{(1)}}-\int_{\boldsymbol{k}}{e\Omega _{ij}^{\ell}E_jf^{(0)}}=\chi _{ij}^{\ell}E_j,
\end{equation}
where $f_{}^{(1)}=\frac{e\tau}{\hbar}\frac{\partial f_{}^{(0)}}{\partial k_j}E_j$. Thus, we can obtain
\begin{equation}
    \chi _{ij}^{\ell}=e\int_{\boldsymbol{k}}{\left[ j_{i}^{\ell}\frac{\tau}{\hbar}\frac{\partial f^{(0)}}{\partial k_j}-\Omega _{ij}^{\ell}f^{(0)} \right]}.
\end{equation}
If we keep to the second order of the electric field, we get
\begin{equation}
    \Gamma _{ijl}^{\ell}=e^2\int_{\boldsymbol{k}}{\left[ j_{i}^{\ell}\frac{\tau ^2}{\hbar ^2}\frac{\partial ^2f^{(0)}}{\partial k_j\partial k_l}-\frac{\tau}{\hbar}\Omega _{ij}^{\ell}\frac{\partial f^{(0)}}{\partial k_l}+\Pi _{ijl}^{\ell}f^{(0)} \right]}.
\end{equation}
\subsection{Symmetry of linear and nonlinear spin currents}
Firstly, what we need to explain is the symmetry of SBC and SBCP. This can be obtained by comparing it with BC.
There is a correlation between them,
\begin{equation}\label{eq}
\begin{aligned}
\Omega _{ij,n}&=2 \mathrm{Im}
\sum_{m\ne n}{\frac{\langle n|\frac{\partial H}{\partial k_i}|m\rangle \langle m|\frac{\partial H}{\partial v_j}|n\rangle}{\left( \varepsilon _{n}^{(0)}-\varepsilon _{m}^{(0)} \right) ^2}}
\\
\Omega _{ij,n}^{\ell} &=2 \mathrm{Im}
\sum_{m\ne n}{\frac{\langle n|\hat{j}_{i}^{\ell}|m\rangle \langle m|\frac{\partial H}{\partial v_j}|n\rangle}{\left( \varepsilon _{n}^{(0)}-\varepsilon _{m}^{(0)} \right) ^2}},\\
\Pi _{ijl}^{\ell} &=\sum_{\substack{m\ne n\\q\ne n}}{\frac{\langle u_{q}^{(0)}|\hat{j}_{i}^{\ell}|u_{m}^{(0)}\rangle \langle u_{n}^{(0)}|\frac{\partial H}{\partial v_j}|u_{q}^{(0)}\rangle \langle u_{m}^{(0)}|\frac{\partial H}{\partial v_l}|u_{n}^{(0)}\rangle}{\left( \varepsilon _{n}^{(0)}-\varepsilon _{q}^{(0)} \right) ^2\left( \varepsilon _{n}^{(0)}-\varepsilon _{m}^{(0)} \right) ^2}}.
\end{aligned}
\end{equation}
Under time reversal $\mathcal{T}$ and space inversion $\mathcal{P}$, the parity of these three terms can be expressed as
\begin{equation}\label{eq}
\begin{aligned}
\mathcal{T} :\quad & \Omega _{ij}\left( k \right) \rightarrow -\Omega _{ij}\left( -k \right),\quad &\Omega _{ij}^{\ell}\left( k \right) \rightarrow \,\,\Omega _{ij}^{\ell}\left( -k \right) ,\quad &\Pi _{ijl}^{\ell}\left( k \right) \rightarrow \,\, \Pi _{ijl}^{\ell}\left( -k \right) ,
\\
\mathcal{P} :\quad & \Omega _{ij}\left( k \right) \rightarrow \Omega _{ij}\left( -k \right) ,\quad&\Omega _{ij}^{\ell}\left( k \right) \rightarrow \,\,\Omega _{ij}^{\ell}\left( -k \right) ,\quad &\Pi _{ijl}^{\ell}\left( k \right) \rightarrow \,\,-\,\,\Pi _{ijl}^{\ell}\left( -k \right) .
\end{aligned}
\end{equation}
For clarity, Table \ref{tableA1} shows all vanishing terms of three classes spin current induced by Drude, SBC and SBCP due to symmetry constants of $\mathcal{T}$ and $\mathcal{P}$.

\begin{table*}[h]
\renewcommand\arraystretch{2}
\centering
\caption{All vanishing spin currents due to symmetry constants of time reversal  $\mathcal{T}$ and space inversion  $\mathcal{P}$.}
\begin{tabular*}{18cm}{@{\extracolsep{\fill}}lccp{5cm}p{5cm}}
\hline\hline
 Contributions     & TRS ($\mathcal{T}$)    & 	IS ($\mathcal{P}$)  \\
\hline
Drude effect ($ p \ge 0$)&  $J_{i,\text{D}}^{\ell \left( p=\text{odd}\right)}=\int_k{j_{i}^{\ell}f^{(p)}}=0$
& $J_{i,D}^{\ell \left( p=\text{even}\right)}=\int_k{j_{i}^{\ell}f^{(p)}}=0$
\\
SBC ($ p \ge1$)  &   \multicolumn{2}{c}{$J_{i,\text{SBC}}^{\ell \left( p=\text{even} \right)}=-e\int_k{\Omega _{ij}^{\ell}E_jf^{(p-1)}}=0$}\\
SBCP ($ p \ge 2$)   &  $ J_{i,\text{SBCP}}^{\ell \left( p=\text{odd} \right)}=e^2\int_k{\Pi _{ijl}^{\ell}E_jE_lf^{(p-2)}}=0$ & $J_{i,\text{SBCP}}^{\ell \left( p=\text{even} \right)}=e^2\int_k{\Pi _{ijl}^{\ell}E_jE_lf^{(p-2)}}=0$ \\
\hline\hline
\end{tabular*}
\label{tableA1}
\end{table*}

\setcounter{figure}{0}
\renewcommand{\thefigure}{B\arabic{figure}}
%\section{The second-order  spin current in the model with SOC}
\section{Spin conductivities in the model with SOC}
\label{appendixB}
By solving the eigen equation, the expression of the wave function is
\begin{equation}
		|+\rangle =\frac{1}{\sqrt{2}}\left[ \begin{array}{c}
			\frac{(\beta k_x-\alpha k_y)-i(\alpha k_x-\beta k_y)}{k\sqrt{\alpha ^2+\beta ^2-4\alpha \beta \cos \phi \sin \phi}}\\
			1\\
		\end{array} \right] ,\quad\quad
			|-\rangle =\frac{1}{\sqrt{2}}\left[ \begin{array}{c}
			-1\\
			\frac{(\beta k_x-\alpha k_y)+i(\alpha k_x-\beta k_y)}{k\sqrt{\alpha ^2+\beta ^2-4\alpha \beta \cos \phi \sin \phi}}\\
		\end{array} \right],
\end{equation}
where $\frac{k_{x}}{k} =\cos\phi, \frac{k_{y}}{k} =\sin \phi$. If we assume that $\displaystyle \varphi =\arg[( \beta k_{x} -\alpha k_{y}) +i( \alpha k_{x} -\beta k_{y})]$, the spin polarization in $k$ space is $\langle u_{\pm } |\vec{\sigma } |u_{\pm } \rangle =\pm (\cos \varphi ,-\sin \varphi ,0) $ shown in Fig. \ref{figs1}. The spin current operators are
\begin{equation}
	\begin{aligned}
		\hat{j}_{x}^x &=\frac{1}{4}\left\{ \frac{\partial H}{\partial k_x},\sigma _x \right\} =\frac{1}{2}\left[ \begin{matrix}
			\beta&		\frac{\hbar^2 k_x}{m}\\
			\frac{\hbar^2  k_x}{m}&		\beta\\
		\end{matrix} \right],
\quad\quad
		\hat{j}_{x}^y=\frac{1}{4}\left\{ \frac{\partial H}{\partial k_x},\sigma _y \right\} =\frac{1}{2}\left[ \begin{matrix}
			\alpha&		-i\frac{\hbar^2  k_x}{m}\\
			i\frac{\hbar^2  k_x}{m}&		\alpha\\
		\end{matrix} \right],
\\
		\hat{j}_{y}^x&=\frac{1}{4}\left\{ \frac{\partial H}{\partial k_y},\sigma _x \right\} =\frac{1}{2}\left[\begin{matrix}
			-\alpha&		\frac{\hbar^2  k_y}{m}\\
			\frac{\hbar^2  k_y}{m}&		-\alpha\\
		\end{matrix} \right],
\quad\quad
		\hat{j}_{y}^y=\frac{1}{4}\left\{ \frac{\partial H}{\partial k_y},\sigma _y \right\} =\frac{1}{2}\left[ \begin{matrix}
			-\beta&		-i\frac{\hbar^2  k_y}{m}\\
			i\frac{\hbar^2  k_y}{m}&		-\beta\\
		\end{matrix} \right].
	\end{aligned}
\end{equation}
\begin{figure*}[t]
	\includegraphics[width=1\linewidth]{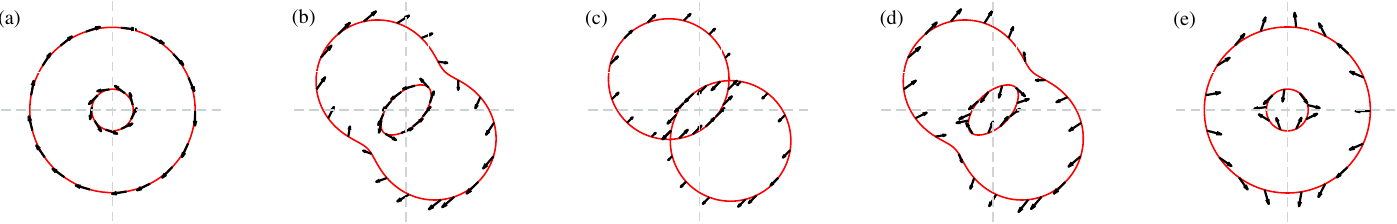}
	\caption{Spin textures of the Fermi surface of the Rashba-Dressselhaus system. (a) For $\theta=0$ (pure Rashba system), (b) For $\theta=\pi/6$, (c) For $\theta=\pi/4$, (d) For $\theta=\pi/3$, (e) For $\theta=\pi/2$ (pure Dresselhaus system).  }
	\label{figs1}
\end{figure*}

Their expectation values for each band are
\begin{equation}\label{eq}
\begin{aligned}
j_{x,\pm}^{x} &=\langle \pm |\hat{j}_{x}^{x}|\pm \rangle =\frac{1}{2}\left( \pm \frac{\hbar ^2k}{m}\cos \phi \cos \varphi +\beta \right) ,\quad \quad j_{x,\pm}^{y}=\langle \pm |\hat{j}_{x}^{y}|\pm \rangle =\frac{1}{2}\left( \mp \frac{\hbar ^2k}{m}\cos \phi \sin \varphi +\alpha \right)
\\
j_{y,\pm}^{x} & =\langle \pm |\hat{j}_{y}^{x}|\pm \rangle =\frac{1}{2}\left( \pm \frac{\hbar ^2k}{m}\sin \phi \cos \varphi -\alpha \right) ,\quad \quad j_{y,\pm}^{y}=\langle \pm |\hat{j}_{y}^{y}|\pm \rangle =\frac{1}{2}\left( \mp \frac{\hbar ^2k}{m}\sin \phi \sin \varphi -\beta \right)
\end{aligned}
\end{equation}
If we suppose the driving field is along the $x$ direction,  we can calculate the Berry curvature polarizability
\begin{equation}
 \Pi _{ixx,\pm}^{\ell}=\left[ \begin{matrix}
	\Pi _{xxx}^{x}&		\Pi _{xxx}^{y}\\
	\Pi _{yxx}^{x}&		\Pi _{yxx}^{y}\\
\end{matrix} \right] =\frac{\left( \alpha ^2-\beta ^2 \right) ^2k_{y}^{2}}{32mk^7A^7(\phi )}\left\{ \mp \hbar ^2\left[ \begin{matrix}
	k_{x}^{2}&		k_xk_y\\
	k_xk_y&		k_{y}^{2}\\
\end{matrix} \right] +mkA\left( \phi \right) \left[ \begin{matrix}
	1&		0\\
	0&		1\\
\end{matrix} \right] \right\} \left[ \begin{matrix}
	\beta&		\alpha\\
	-\alpha&		-\beta\\
\end{matrix} \right] ,
\end{equation}
where
\begin{equation}\label{eqB5}
\begin{aligned}
\Pi _{xxx,\pm}^{x} &=\frac{\left( \alpha ^2-\beta ^2 \right) ^2k_{y}^{2}\left[ \mp \beta \hbar ^2k_{x}^{2}\pm \alpha \hbar ^2k_xk_y+\beta m\sqrt{\left( \alpha ^2+\beta ^2 \right) k^2-4\alpha \beta k_xk_y} \right]}{32m\left[ \sqrt{\left( \alpha ^2+\beta ^2 \right) k^2-4\alpha \beta k_xk_y} \right] ^7},
\\
\Pi _{xxx,\pm}^{y} &=\frac{\left( \alpha ^2-\beta ^2 \right) ^2k_{y}^{2}\left[ \mp \alpha \hbar ^2k_{x}^{2}\pm \beta \hbar ^2k_xk_y+\alpha m\sqrt{\left( \alpha ^2+\beta ^2 \right) k^2-4\alpha \beta k_xk_y} \right]}{32m\left[ \sqrt{\left( \alpha ^2+\beta ^2 \right) k^2-4\alpha \beta k_xk_y} \right] ^7},
\\
\Pi _{yxx,\pm}^{x} &=\frac{\left( \alpha ^2-\beta ^2 \right) ^2k_{y}^{2}\left[ \mp \beta \hbar ^2k_xk_y\pm \alpha \hbar ^2k_{y}^{2}-\alpha m\sqrt{\left( \alpha ^2+\beta ^2 \right) k^2-4\alpha \beta k_xk_y} \right]}{32m\left[ \sqrt{\left( \alpha ^2+\beta ^2 \right) k^2-4\alpha \beta k_xk_y} \right] ^7},
\\
\Pi _{yxx,\pm}^{y} &=\frac{\left( \alpha ^2-\beta ^2 \right) ^2k_{y}^{2}\left[ \mp \alpha \hbar ^2k_xk_y\pm \beta \hbar ^2k_{y}^{2}-\beta m\sqrt{\left( \alpha ^2+\beta ^2 \right) k^2-4\alpha \beta k_xk_y} \right]}{32m\left[ \sqrt{\left( \alpha ^2+\beta ^2 \right) k^2-4\alpha \beta k_xk_y} \right] ^7}.
\end{aligned}
\end{equation}

\subsection{The zeroth spin conductivity in thermodynamic equilibrium}
\label{appendixB1}
According to Eq. (\ref{eqA11}), the zero-order spin conductivity in thermodynamic equilibrium can be expressed as
\begin{equation}
    \sigma _{i}^{\ell}=\sum_{\pm}^{}{\int_k^{}{j_{i,\pm}^{\ell}f_0}}.
\end{equation}
Here we take the pure Rashba system ($\alpha=1, \beta=0$) as an example. When the chemical potential is positive, $\mu >0$,
\begin{equation}\label{eq}
\begin{aligned}
\sigma _{x}^{x} &=\sum_{\pm}^{}{\int_k^{}{j_{x,\pm}^{x}f_0}}=\sum_{\pm}^{}{\int{\frac{d^2k}{\left( 2\pi \right) ^2}\mp \frac{\hbar ^2k_xk_y}{2m}f_0=0}},
\\
\sigma _{y}^{y} &=\sum_{\pm}^{}{\int_k^{}{j_{y,\pm}^{y}f_0}}=\sum_{\pm}^{}{\int{\frac{d^2k}{\left( 2\pi \right) ^2}\pm \frac{\hbar ^2k_xk_y}{2m}f_0=0}},
\end{aligned}
\end{equation}
and
\begin{equation}\label{eq}
\begin{aligned}
\sigma _{x}^{y} &=\sum_{\pm}^{}{\int_k^{}{j_{x,\pm}^{y}f_0}}=\sum_{\pm}^{}{\int{\frac{d^2k}{\left( 2\pi \right) ^2}\left( \pm \frac{\hbar ^2k_{x}^{2}}{2mk}+\frac{\alpha}{2} \right) f_0}}=\frac{\hbar ^2}{8m\pi ^2}\sum_{\pm}^{}{\int_0^{k_F}{\pm \frac{k^2\cos ^2\phi}{k}kdkd\phi}}+\frac{\alpha}{8\pi ^2}\sum_{\pm}^{}{\int_0^{k_F}{kdkd\phi}}
\\
&=\frac{\hbar ^2}{24m\pi}\left[ \left( k_{F^+}^{+} \right) ^3-\left( k_{F^-}^{+} \right) ^3 \right] +\frac{\alpha}{8\pi}\left[ \left( k_{F^+}^{+} \right) ^2-\left( k_{F^-}^{+} \right) ^2 \right] =\frac{\alpha ^3m^3}{6\hbar ^4\pi},
\\
\sigma _{y}^{x}
&=\sum_{\pm}^{}{\int_k^{}{j_{y,\pm}^{x}f_0}}=\sum_{\pm}^{}{\int{\frac{d^2k}{\left( 2\pi \right) ^2}\left( \mp \frac{\hbar ^2k_{y}^{2}}{2mk}-\frac{\alpha}{2} \right) f_0}}=-\sigma _{x}^{y}=-\frac{\alpha ^3m^3}{6\hbar ^4\pi},
\end{aligned}
\end{equation}
which  is equal to the result in Ref. \cite{Rashba2003} multiplied by $\frac{\hbar}{2}$ since we are taking the spin $s^{x/y}=\frac{\hbar}{2} \sigma^{x/y}$.

When $\alpha=0, \beta =1$, the four components of zeroth spin conductivity are
\begin{equation}
    \sigma _{x}^{x}=-\sigma _{y}^{y}=\frac{\beta ^3m^3}{6\hbar ^4\pi},\quad\quad \sigma _{x}^{y}=\sigma _{y}^{x}=0.
\end{equation}
In other cases, we can get the corresponding result according to numerical calculation.

\subsection{The extrinsic term of second order spin conductivity}
\label{appendixB2}
It turns out that  the partial derivative of the distribution function with respect to energy is the delta function if we approach the zero temperature limit. We further consider the properties of the delta function,
\begin{equation}
	\begin{aligned}
		&\frac{\partial f_0}{\partial \varepsilon ^+}=-\delta \left( \varepsilon ^+-\mu \right) =-\frac{\delta \left(k-k_{F_-}^{+} \right)}{\left|\frac{\partial \varepsilon ^+}{\partial k_{F^+}^{+}}\right|}\left( \mu >0 \right),
		\\
		&\frac{\partial f_0}{\partial \varepsilon ^-}=-\delta \left( \varepsilon ^--\mu \right) =-\frac{\delta \left( k-k_{F_-}^{+} \right)}{\left|\frac{\partial \varepsilon ^-}{\partial k_{F^-}^{+}}\right|}\left( \mu >0 \right),
		\\
		&\frac{\partial f_0}{\partial \varepsilon ^-}=-\left[ \frac{\delta \left( k-k_{F_-}^{+} \right)}{\left|\frac{\partial \varepsilon ^+}{\partial k_{F^-}^{+}}\right|}+\frac{\delta \left( k-k_{F_-}^{-} \right)}{|\frac{\partial \varepsilon ^+}{\partial k_{F^-}^{-}}|} \right] \left( \mu <0 \right).
	\end{aligned}
\end{equation}
Next, we will use these equations to obtain the second-order spin current.

Here we take the $+$ band as an example,
\begin{equation}
	\begin{aligned}
		\Gamma_{xxx,+}^{y, \text{ext}}&=\int{\frac{d^2k}{\left( 2\pi \right) ^2}\langle u_+|}	\hat{j}_{x}^y |u_+\rangle f_{2}^{+}
=\int{\frac{d^2k}{\left( 2\pi \right) ^2}\frac{1}{2}\left( -\frac{\hbar^2  k}{m}\frac{k_x}{k}\frac{\beta k_y-\alpha k_x}{kA\left( \phi \right)}+\alpha \right)}\left( \frac{e\tau}{\hbar} \right) ^2\frac{\partial ^2f_0}{\partial k_{x}^{2}}
		\\
		&=\frac{e^2\tau ^2}{2\hbar^2 \left( 2\pi \right) ^2}\int{d^2k\left( -\frac{\hbar^2  k}{m}\frac{k_x}{k}\frac{\beta k_y-\alpha k_x}{kA\left( \phi \right)}+\alpha \right) \frac{\partial}{\partial k_x}\left[ \left( \frac{\left( \alpha ^2+\beta ^2 \right) k_x-2\alpha \beta k_y}{kA\left( \phi \right)}+\frac{\hbar^2 k_x}{m} \right) \frac{\partial f_0}{\partial \varepsilon ^+} \right]}
		\\
		 &=\frac{e^2\tau ^2}{2 \hbar^2  \left( 2\pi \right) ^2}\int{d^2k\frac{\partial}{\partial k_x}\left( -\frac{\hbar^2  k}{m}\frac{k_x}{k}\frac{\beta k_y-\alpha k_x}{kA\left( \phi \right)}+\alpha \right) \left[ \left( \frac{\left( \alpha ^2+\beta ^2 \right) k_x-2\alpha \beta k_y}{kA\left( \phi \right)}+\frac{\hbar^2 k_x}{m} \right) \delta \left( \varepsilon ^+-\mu \right) \right]},
	\end{aligned}
\end{equation}
where
\begin{equation}
	\frac{\partial}{\partial k_x}\left( -\frac{\hbar^2  k}{m}\frac{k_x}{k}\frac{\beta k_y-\alpha k_x}{kA\left( \phi \right)}+\alpha \right) = \frac{\hbar^2 }{m}\frac{\alpha (\alpha ^2+\beta^2)k_{x}^{3}-6\alpha ^2\beta k_{x}^{2}k_y+2\alpha (\alpha ^2+2\beta ^2)k_xk_{y}^{2}-\beta(\alpha ^2+\beta ^2)k_{y}^{3}}{k^3A\left( \phi \right) ^3}.
\end{equation}
Thus,
\begin{equation}
	\begin{aligned}
		\Gamma_{xxx,+}^{y, \text{ext}}&=\frac{e^2\tau ^2}{2m\left( 2\pi \right) ^2}\int{d\phi \frac{k_{F^+}^{+}}{\left|\frac{\hbar^2 k^{+}_{F^+}}{m}+A\left( \phi \right) \right|}\left( \frac{\left( \alpha ^2+\beta ^2 \right) \cos \phi -2\alpha \beta \sin \phi}{A\left( \phi \right)}+\frac{\hbar^2 k_{F^+}^{+} \cos \phi}{m} \right) \,\,   }
		\\
		&\times \left[ \frac{\alpha \left( \alpha ^2+\beta ^2 \right) \cos ^3\phi}{A\left( \phi \right) ^3}-\frac{6\alpha ^2\beta \cos ^2\phi \sin \phi}{A\left( \phi \right) ^3}+\frac{2\alpha (\alpha ^2+2\beta ^2)\cos \phi \sin ^2\phi}{A\left( \phi \right) ^3}-\frac{\beta (\alpha ^2+\beta ^2)\sin ^3\phi}{A\left( \phi \right) ^3} \right].
	\end{aligned}
\end{equation}
and

And the other three spin conductivities are
\begin{equation}
\begin{aligned}
   \Gamma_{xxx,+}^{x, \text{ext}}&=\frac{e^{2} \tau ^{2} }{8m\pi ^{2}}\int d\phi \left(\frac{\hbar^2  k_{F+}^{+}\cos \phi }{m} +\frac{\left( \alpha ^{2} +\beta ^{2}\right)\cos \phi -2\alpha \beta \sin \phi }{A( \phi )}\right)\frac{k_{F+}^{+}}{\left|\frac{\hbar^2  k_{F+}^{+}}{m} +A( \phi )\right|}\\
	&\times  \left[\frac{\beta \left( \alpha ^{2} +\beta ^{2}\right)\cos^{3} \phi -6\alpha \beta ^{2}\cos^{2} \phi \sin \phi +2\beta \left( 2\alpha ^{2} +\beta ^{2}\right)\cos \phi \sin^{2} \phi -\alpha \left( \alpha ^{2} +\beta ^{2}\right)\sin^{3} \phi }{A( \phi )^{3}}\right],\\
	\Gamma_{yxx,+}^{x, \text{ext}}&=\frac{e^{2} \tau ^{2} }{8m\pi ^{2}}\int d\phi \left(\frac{\hbar^2  k_{F+}^{+}\cos \phi }{m} +\frac{\left( \alpha ^{2} +\beta ^{2}\right)\cos \phi -2\alpha \beta \sin \phi }{A( \phi )}\right)\frac{k_{F+}^{+}}{\left|\frac{\hbar^2  k_{F+}^{+}}{m} +A( \phi )\right|}\left[\frac{\left( \alpha ^{2} -\beta ^{2}\right) k_{y}^{2}( \alpha k_{x} -\beta k_{y})}{A^{3}}\right],\\
		\Gamma_{yxx,+}^{y, \text{ext}}& =\frac{e^{2} \tau ^{2} }{8m \pi ^{2}}\int d\phi \left(\frac{\hbar^2  k_{F+}^{+}\cos \phi }{m} +\frac{\left( \alpha ^{2} +\beta ^{2}\right)\cos \phi -2\alpha \beta \sin \phi }{A( \phi )}\right)\frac{k_{F+}^{+}}{\left|\frac{\hbar^2  k_{F+}^{+}}{m} +A( \phi ) \right|}\left[\frac{\left(\alpha ^{2}-\beta ^{2}\right) k_{y}^{2}( \alpha k_{y} -\beta k_{x})}{A^{3}}\right].
	\end{aligned}
\end{equation}

\subsubsection{Pure Rashba system}
A simple case describes the pure Rashba system with $\alpha=1,\beta=0$, which does not allow the longitudinal intrinsic second-order spin current.
%\begin{equation}
%	H_{RD}=\frac{\hbar^2 k^2}{2m}+\alpha (k_x\sigma^y-k_y\sigma^x).
%\end{equation}
We can get
\begin{equation}
	\begin{aligned}
		\Gamma_{xxx}^{y, \text{ext}} &=\sum_{\pm}^{}{\frac{e^2\tau ^2}{2\hbar ^2\left( 2\pi \right) ^2}\int{kdkd\phi \left[ \pm \frac{\hbar ^2k_{x}^{2}}{m}+\alpha \right]}\frac{\partial}{\partial k_x}\left[ \left( \frac{\hbar ^2k_x}{m}\pm \frac{\alpha k_x}{k} \right) \frac{\partial f_0}{\partial \varepsilon ^\pm} \right]}
\\
&=\sum_{\pm}^{}{\frac{e^2\tau ^2}{2m\left( 2\pi \right) ^2}\int{kdkd\phi}\left[ \pm \frac{k_{x}^{3}+2k_xk_{y}^{2}}{k^3} \right] \left[ \left( \frac{\hbar ^2k}{m}\pm \alpha \right) \cos \phi \delta \left( \varepsilon ^\pm-\mu \right) \right]}
\\
&=\frac{5e^2\tau ^2}{32\pi m}\begin{cases}
	k_{F^+}^{+}-k_{F^-}^{+} \,\,\left( \mu >0 \right)\\
	-k_{F^-}^{+}+k_{F^-}^{-}\,\,\left( \mu <0 \right)\\
\end{cases}
		\\
		\Gamma _{yxx}^{x,\text{ext}} &=\sum_{\pm}^{}{\frac{e^2\tau ^2}{2\hbar ^2\left( 2\pi \right) ^2}\int{kdkd\phi \left[ \mp \frac{\hbar ^2k_{y}^{2}}{m}-\alpha \right]}\frac{\partial}{\partial k_x}\left[ \left( \frac{\hbar ^2k_x}{m}\pm \frac{\alpha k_x}{k} \right) \frac{\partial f_0}{\partial \varepsilon ^\pm} \right]}
\\
&=\sum_{\pm}^{}{\frac{e^2\tau ^2}{2\hbar ^2\left( 2\pi \right) ^2}\int{kdkd\phi}\left[ \pm \frac{\hbar ^2k_xk_{y}^{2}}{mk^3} \right] \left[ \left( \frac{\hbar ^2k}{m}\pm \alpha \right) \cos \phi \delta \left( \varepsilon ^\pm-\mu \right) \right]}
\\
&=\frac{e^2\tau ^2}{32\pi m}\begin{cases}
	k_{F^+}^{+}-k_{F^-}^{+}\,\,\left( \mu >0 \right)\\
	-k_{F^-}^{+}+k_{F^-}^{-}\,\,\left( \mu <0 \right)\\
\end{cases}=\frac{1}{5}\Gamma _{xxx}^{y,\text{ext}},
	\end{aligned}
\end{equation}
where $	\hbar^2 k_{F^+}^{+} =-m\alpha +\sqrt{m^2\alpha ^2+2\hbar^2m\mu}$, $\hbar^2 k_{F^-}^{\pm} =m\alpha \pm \sqrt{m^2\alpha ^2+2\hbar^2 m\mu}$. Here we set the relaxation time $\tau \approx 0.1$ ps and  the unit of spin conductivity is $\mathrm{e}\cdot \mathrm{\mu m}/\mathrm{V}$.
The other terms are all zeros due to the angle integration.
\begin{equation}\label{eq}
\begin{aligned}
\Gamma _{xxx}^{x, \text{ext}} &=\sum_{\pm} \frac{e^2\tau ^2}{2\hbar ^2\left( 2\pi \right) ^2}\int{kdkd\phi \left[ \mp \frac{\hbar ^2k_xk_y}{mk} \right]}\frac{\partial}{\partial k_x}\left[ \left( \frac{\hbar ^2k_x}{m}\pm \frac{\alpha k_x}{k} \right) \frac{\partial f_0}{\partial \varepsilon ^\pm} \right]
\\
&=\sum_{\pm} \mp \frac{e^2\tau ^2}{2m\left( 2\pi \right) ^2}\int{kdkd\phi}\left[ \cos \phi \sin ^3\phi \right] \left[ \left( \frac{\hbar ^2k}{m}\pm \alpha \right) \delta \left( \varepsilon ^\pm-\mu \right) \right] =0.
\\
\Gamma _{yxx}^{y, \text{ext}} &=\sum_{\pm} \frac{e^2\tau ^2}{2\hbar ^2\left( 2\pi \right) ^2}\int{kdkd\phi \left[ \pm \frac{\hbar ^2k_xk_y}{mk} \right]}\frac{\partial}{\partial k_x}\left[ \left( \frac{\hbar ^2k_x}{m}\pm \frac{\alpha k_x}{k} \right) \frac{\partial f_0}{\partial \varepsilon ^\pm} \right] =0.
\end{aligned}
\end{equation}

\subsubsection{Pure Dresselhaus system}
For the pure Dressselhaus system with $\alpha=0,\beta=1$, we can use the same computational idea to get
\begin{equation}
	\begin{aligned}
		\Gamma _{xxx}^{x,\mathrm{ext}} &=\sum_{\pm}^{}{\frac{e^2\tau ^2}{2\hbar ^2\left( 2\pi \right) ^2}\int{kdkd\phi \left[ \pm \frac{\hbar ^2k_{x}^{2}}{m}+\beta \right]}\frac{\partial}{\partial k_x}\left[ \left( \frac{\hbar ^2k_x}{m}\pm \frac{\beta k_x}{k} \right) \frac{\partial f_0}{\partial \varepsilon ^{\pm}} \right]}
\\
&=\frac{5e^2\tau ^2}{32\pi m}\begin{cases}
	k_{F^+}^{+}-k_{F^-}^{+}\left( \mu >0 \right)\\
	-k_{F^-}^{+}+k_{F^-}^{-}\,\,\left( \mu <0 \right)\\
\end{cases}
		\\
		\Gamma_{yxx}^{y, \text{ext}}&=\frac{1}{5}\Gamma_{xxx}^{x, \text{ext}}.\\
	\end{aligned}
	\label{eqb11}
\end{equation}

\subsection{The intrinsic term of second order spin conductivity}
According to Eq. (\ref{eqB5}), the second order intrinsic spin conductivity can be obtained by integrating in the $k$ space. Here we take $\Gamma _{xxx}^{x,\mathrm{int}}$ as an example,
\begin{equation}\label{eq}
\begin{aligned}
\Gamma _{xxx}^{x,\mathrm{int}} &=e^2\sum_{\pm}^{}{\int{kdkd\phi}\frac{\left( \alpha ^2-\beta ^2 \right) ^2k_{y}^{2}\left[ \mp \beta \hbar ^2k_{x}^{2}\pm \alpha \hbar ^2k_xk_y+\beta m\sqrt{\left( \alpha ^2+\beta ^2 \right) k^2-4\alpha \beta k_xk_y} \right]}{32m\left[ \sqrt{\left( \alpha ^2+\beta ^2 \right) k^2-4\alpha \beta k_xk_y} \right] ^7}f^{(0)}.}
\end{aligned}
\end{equation}
When $\alpha =1, \beta =0$,
\begin{equation}
    \Gamma _{xxx}^{x,\mathrm{int}}=e^2\sum_{\pm}^{}{\int{kdkd\phi}\frac{ k_{y}^{2}\left[ \pm \alpha \hbar ^2k_xk_y \right]}{32m \alpha^3 k^7}f^{(0)}=0}.
\end{equation}
When $\alpha=0, \beta=1$,
\begin{equation}
    \Gamma _{xxx}^{x,\mathrm{int}}=e^2\sum_{\pm}^{}{\int{kdkd\phi}\frac{k_{y}^{2}\left[ \mp \beta \hbar ^2k_{x}^{2}+\beta ^2km \right]}{32m\beta^3k^7}f^{(0)}}.
\end{equation}
where $f^{(0)}$  is the Fermi-Dirac distribution function and we take the temperature $T=10$ K.
\end{widetext}

%\bibliography{ref}

\begin{thebibliography}{46}%
\makeatletter
\providecommand \@ifxundefined [1]{%
 \@ifx{#1\undefined}
}%
\providecommand \@ifnum [1]{%
 \ifnum #1\expandafter \@firstoftwo
 \else \expandafter \@secondoftwo
 \fi
}%
\providecommand \@ifx [1]{%
 \ifx #1\expandafter \@firstoftwo
 \else \expandafter \@secondoftwo
 \fi
}%
\providecommand \natexlab [1]{#1}%
\providecommand \enquote  [1]{``#1''}%
\providecommand \bibnamefont  [1]{#1}%
\providecommand \bibfnamefont [1]{#1}%
\providecommand \citenamefont [1]{#1}%
\providecommand \href@noop [0]{\@secondoftwo}%
\providecommand \href [0]{\begingroup \@sanitize@url \@href}%
\providecommand \@href[1]{\@@startlink{#1}\@@href}%
\providecommand \@@href[1]{\endgroup#1\@@endlink}%
\providecommand \@sanitize@url [0]{\catcode `\\12\catcode `\$12\catcode
  `\&12\catcode `\#12\catcode `\^12\catcode `\_12\catcode `\%12\relax}%
\providecommand \@@startlink[1]{}%
\providecommand \@@endlink[0]{}%
\providecommand \url  [0]{\begingroup\@sanitize@url \@url }%
\providecommand \@url [1]{\endgroup\@href {#1}{\urlprefix }}%
\providecommand \urlprefix  [0]{URL }%
\providecommand \Eprint [0]{\href }%
\providecommand \doibase [0]{https://doi.org/}%
\providecommand \selectlanguage [0]{\@gobble}%
\providecommand \bibinfo  [0]{\@secondoftwo}%
\providecommand \bibfield  [0]{\@secondoftwo}%
\providecommand \translation [1]{[#1]}%
\providecommand \BibitemOpen [0]{}%
\providecommand \bibitemStop [0]{}%
\providecommand \bibitemNoStop [0]{.\EOS\space}%
\providecommand \EOS [0]{\spacefactor3000\relax}%
\providecommand \BibitemShut  [1]{\csname bibitem#1\endcsname}%
\let\auto@bib@innerbib\@empty
%</preamble>
\bibitem [{\citenamefont {v.~Klitzing}\ \emph {et~al.}(1980)\citenamefont
  {v.~Klitzing}, \citenamefont {Dorda},\ and\ \citenamefont
  {Pepper}}]{Klitzing}%
  \BibitemOpen
  \bibfield  {author} {\bibinfo {author} {\bibfnamefont {K.}~\bibnamefont
  {v.~Klitzing}}, \bibinfo {author} {\bibfnamefont {G.}~\bibnamefont {Dorda}},\
  and\ \bibinfo {author} {\bibfnamefont {M.}~\bibnamefont {Pepper}},\
  }\bibfield  {title} {\bibinfo {title} {New method for high-accuracy
  determination of the fine-structure constant based on quantized hall
  resistance},\ }\href {https://doi.org/10.1103/physrevlett.45.494} {\bibfield
  {journal} {\bibinfo  {journal} {Phys. Rev. Lett.}\ }\textbf {\bibinfo
  {volume} {45}},\ \bibinfo {pages} {494} (\bibinfo {year} {1980})}\BibitemShut
  {NoStop}%
\bibitem [{\citenamefont {Cage}\ \emph {et~al.}(1989)\citenamefont {Cage},
  \citenamefont {Klitzing}, \citenamefont {Chang}, \citenamefont {Duncan},
  \citenamefont {Haldane}, \citenamefont {Laughlin}, \citenamefont {Pruisken},\
  and\ \citenamefont {Thouless}}]{Prange}%
  \BibitemOpen
  \bibfield  {author} {\bibinfo {author} {\bibfnamefont {M.~E.}\ \bibnamefont
  {Cage}}, \bibinfo {author} {\bibfnamefont {K.}~\bibnamefont {Klitzing}},
  \bibinfo {author} {\bibfnamefont {A.}~\bibnamefont {Chang}}, \bibinfo
  {author} {\bibfnamefont {F.}~\bibnamefont {Duncan}}, \bibinfo {author}
  {\bibfnamefont {M.}~\bibnamefont {Haldane}}, \bibinfo {author} {\bibfnamefont
  {R.}~\bibnamefont {Laughlin}}, \bibinfo {author} {\bibfnamefont
  {A.}~\bibnamefont {Pruisken}},\ and\ \bibinfo {author} {\bibfnamefont
  {D.}~\bibnamefont {Thouless}},\ }\href@noop {} {\emph {\bibinfo {title} {The
  Quantum Hall Effect}}}\ (\bibinfo  {publisher} {Springer, New York},\
  \bibinfo {year} {1989})\BibitemShut {NoStop}%
\bibitem [{\citenamefont {Xiao}\ \emph {et~al.}(2010)\citenamefont {Xiao},
  \citenamefont {Chang},\ and\ \citenamefont {Niu}}]{Xiao}%
  \BibitemOpen
  \bibfield  {author} {\bibinfo {author} {\bibfnamefont {D.}~\bibnamefont
  {Xiao}}, \bibinfo {author} {\bibfnamefont {M.-C.}\ \bibnamefont {Chang}},\
  and\ \bibinfo {author} {\bibfnamefont {Q.}~\bibnamefont {Niu}},\ }\bibfield
  {title} {\bibinfo {title} {Berry phase effects on electronic properties},\
  }\href {https://doi.org/10.1103/revmodphys.82.1959} {\bibfield  {journal}
  {\bibinfo  {journal} {Rev. Mod. Phys.}\ }\textbf {\bibinfo {volume} {82}},\
  \bibinfo {pages} {1959} (\bibinfo {year} {2010})}\BibitemShut {NoStop}%
\bibitem [{\citenamefont {Onsager}(1931{\natexlab{a}})}]{Onsager1931}%
  \BibitemOpen
  \bibfield  {author} {\bibinfo {author} {\bibfnamefont {L.}~\bibnamefont
  {Onsager}},\ }\bibfield  {title} {\bibinfo {title} {Reciprocal relations in
  irreversible processes. i.},\ }\href {https://doi.org/10.1103/physrev.37.405}
  {\bibfield  {journal} {\bibinfo  {journal} {Phys. Rev.}\ }\textbf {\bibinfo
  {volume} {37}},\ \bibinfo {pages} {405} (\bibinfo {year}
  {1931}{\natexlab{a}})}\BibitemShut {NoStop}%
\bibitem [{\citenamefont {Onsager}(1931{\natexlab{b}})}]{Onsager1931a}%
  \BibitemOpen
  \bibfield  {author} {\bibinfo {author} {\bibfnamefont {L.}~\bibnamefont
  {Onsager}},\ }\bibfield  {title} {\bibinfo {title} {Reciprocal relations in
  irreversible processes. ii.},\ }\href
  {https://doi.org/10.1103/physrev.38.2265} {\bibfield  {journal} {\bibinfo
  {journal} {Phys. Rev.}\ }\textbf {\bibinfo {volume} {38}},\ \bibinfo {pages}
  {2265} (\bibinfo {year} {1931}{\natexlab{b}})}\BibitemShut {NoStop}%
\bibitem [{\citenamefont {Wang}\ \emph {et~al.}(2021)\citenamefont {Wang},
  \citenamefont {Gao},\ and\ \citenamefont {Xiao}}]{Wang2021}%
  \BibitemOpen
  \bibfield  {author} {\bibinfo {author} {\bibfnamefont {C.}~\bibnamefont
  {Wang}}, \bibinfo {author} {\bibfnamefont {Y.}~\bibnamefont {Gao}},\ and\
  \bibinfo {author} {\bibfnamefont {D.}~\bibnamefont {Xiao}},\ }\bibfield
  {title} {\bibinfo {title} {Intrinsic nonlinear hall effect in
  antiferromagnetic tetragonal {CuMnAs}},\ }\href
  {https://doi.org/10.1103/physrevlett.127.277201} {\bibfield  {journal}
  {\bibinfo  {journal} {Phys. Rev. Lett.}\ }\textbf {\bibinfo {volume} {127}},\
  \bibinfo {pages} {277201} (\bibinfo {year} {2021})}\BibitemShut {NoStop}%
\bibitem [{\citenamefont {Zhang}\ \emph {et~al.}(2023)\citenamefont {Zhang},
  \citenamefont {Zhu},\ and\ \citenamefont {Su}}]{Zhang2023}%
  \BibitemOpen
  \bibfield  {author} {\bibinfo {author} {\bibfnamefont {Z.-F.}\ \bibnamefont
  {Zhang}}, \bibinfo {author} {\bibfnamefont {Z.-G.}\ \bibnamefont {Zhu}},\
  and\ \bibinfo {author} {\bibfnamefont {G.}~\bibnamefont {Su}},\ }\bibfield
  {title} {\bibinfo {title} {Symmetry dictionary on charge and spin nonlinear
  responses for all magnetic point groups with nontrivial topological nature},\
  }\href {https://doi.org/10.1093/nsr/nwad104} {\bibfield  {journal} {\bibinfo
  {journal} {Natl. Sci. Rev.}\ }\textbf {\bibinfo {volume} {10}},\ \bibinfo
  {pages} {nwad104} (\bibinfo {year} {2023})}\BibitemShut {NoStop}%
\bibitem [{\citenamefont {Sodemann}\ and\ \citenamefont
  {Fu}(2015)}]{Sodemann2015}%
  \BibitemOpen
  \bibfield  {author} {\bibinfo {author} {\bibfnamefont {I.}~\bibnamefont
  {Sodemann}}\ and\ \bibinfo {author} {\bibfnamefont {L.}~\bibnamefont {Fu}},\
  }\bibfield  {title} {\bibinfo {title} {Quantum nonlinear hall effect induced
  by berry curvature dipole in time-reversal invariant materials},\ }\href
  {https://doi.org/10.1103/physrevlett.115.216806} {\bibfield  {journal}
  {\bibinfo  {journal} {Phys. Rev. Lett.}\ }\textbf {\bibinfo {volume} {115}},\
  \bibinfo {pages} {216806} (\bibinfo {year} {2015})}\BibitemShut {NoStop}%
\bibitem [{\citenamefont {Du}\ \emph {et~al.}(2018)\citenamefont {Du},
  \citenamefont {Wang}, \citenamefont {Lu},\ and\ \citenamefont
  {Xie}}]{Du2018}%
  \BibitemOpen
  \bibfield  {author} {\bibinfo {author} {\bibfnamefont {Z.~Z.}\ \bibnamefont
  {Du}}, \bibinfo {author} {\bibfnamefont {C.~M.}\ \bibnamefont {Wang}},
  \bibinfo {author} {\bibfnamefont {H.-Z.}\ \bibnamefont {Lu}},\ and\ \bibinfo
  {author} {\bibfnamefont {X.~C.}\ \bibnamefont {Xie}},\ }\bibfield  {title}
  {\bibinfo {title} {Band signatures for strong nonlinear hall effect in
  bilayer {WTe}$_2$},\ }\href {https://doi.org/10.1103/physrevlett.121.266601}
  {\bibfield  {journal} {\bibinfo  {journal} {Phys. Rev. Lett.}\ }\textbf
  {\bibinfo {volume} {121}},\ \bibinfo {pages} {266601} (\bibinfo {year}
  {2018})}\BibitemShut {NoStop}%
\bibitem [{\citenamefont {Ma}\ \emph {et~al.}(2019)\citenamefont {Ma},
  \citenamefont {Xu}, \citenamefont {Shen}, \citenamefont {MacNeill},
  \citenamefont {Fatemi},\ and\ \citenamefont {\emph{et al.}}}]{Qiong}%
  \BibitemOpen
  \bibfield  {author} {\bibinfo {author} {\bibfnamefont {Q.}~\bibnamefont
  {Ma}}, \bibinfo {author} {\bibfnamefont {S.-Y.}\ \bibnamefont {Xu}}, \bibinfo
  {author} {\bibfnamefont {H.}~\bibnamefont {Shen}}, \bibinfo {author}
  {\bibfnamefont {D.}~\bibnamefont {MacNeill}}, \bibinfo {author}
  {\bibfnamefont {V.}~\bibnamefont {Fatemi}},\ and\ \bibinfo {author}
  {\bibnamefont {\emph{et al.}}},\ }\bibfield  {title} {\bibinfo {title}
  {Observation of the nonlinear hall effect under time-reversal-symmetric
  conditions},\ }\href {https://doi.org/10.1038/s41586-018-0807-6} {\bibfield
  {journal} {\bibinfo  {journal} {Nature}\ }\textbf {\bibinfo {volume} {565}},\
  \bibinfo {pages} {337} (\bibinfo {year} {2019})}\BibitemShut {NoStop}%
\bibitem [{\citenamefont {Kang}\ \emph {et~al.}(2019)\citenamefont {Kang},
  \citenamefont {Li}, \citenamefont {Sohn}, \citenamefont {Shan},\ and\
  \citenamefont {Mak}}]{Kang}%
  \BibitemOpen
  \bibfield  {author} {\bibinfo {author} {\bibfnamefont {K.}~\bibnamefont
  {Kang}}, \bibinfo {author} {\bibfnamefont {T.}~\bibnamefont {Li}}, \bibinfo
  {author} {\bibfnamefont {E.}~\bibnamefont {Sohn}}, \bibinfo {author}
  {\bibfnamefont {J.}~\bibnamefont {Shan}},\ and\ \bibinfo {author}
  {\bibfnamefont {K.~F.}\ \bibnamefont {Mak}},\ }\bibfield  {title} {\bibinfo
  {title} {Nonlinear anomalous hall effect in few-layer {WTe}$_2$},\ }\href
  {https://doi.org/10.1038/s41563-019-0294-7} {\bibfield  {journal} {\bibinfo
  {journal} {Nat. Mater.}\ }\textbf {\bibinfo {volume} {18}},\ \bibinfo {pages}
  {324} (\bibinfo {year} {2019})}\BibitemShut {NoStop}%
\bibitem [{\citenamefont {Li}\ \emph {et~al.}(2023)\citenamefont {Li},
  \citenamefont {Li}, \citenamefont {Xiao}, \citenamefont {Liu}, \citenamefont
  {Wu}, \citenamefont {Gan}, \citenamefont {Han}, \citenamefont {Tang},
  \citenamefont {Zhang},\ and\ \citenamefont {Wang}}]{Li2023}%
  \BibitemOpen
  \bibfield  {author} {\bibinfo {author} {\bibfnamefont {H.}~\bibnamefont
  {Li}}, \bibinfo {author} {\bibfnamefont {M.}~\bibnamefont {Li}}, \bibinfo
  {author} {\bibfnamefont {R.-C.}\ \bibnamefont {Xiao}}, \bibinfo {author}
  {\bibfnamefont {W.}~\bibnamefont {Liu}}, \bibinfo {author} {\bibfnamefont
  {L.}~\bibnamefont {Wu}}, \bibinfo {author} {\bibfnamefont {W.}~\bibnamefont
  {Gan}}, \bibinfo {author} {\bibfnamefont {H.}~\bibnamefont {Han}}, \bibinfo
  {author} {\bibfnamefont {X.}~\bibnamefont {Tang}}, \bibinfo {author}
  {\bibfnamefont {C.}~\bibnamefont {Zhang}},\ and\ \bibinfo {author}
  {\bibfnamefont {J.}~\bibnamefont {Wang}},\ }\bibfield  {title} {\bibinfo
  {title} {Current induced second-order nonlinear hall effect in bulk
  {WTe}$_2$},\ }\href {https://doi.org/10.1063/5.0172026} {\bibfield  {journal}
  {\bibinfo  {journal} {Appl. Phys. Lett.}\ }\textbf {\bibinfo {volume}
  {123}},\ \bibinfo {pages} {163102} (\bibinfo {year} {2023})}\BibitemShut
  {NoStop}%
\bibitem [{\citenamefont {Ye}\ \emph {et~al.}(2023)\citenamefont {Ye},
  \citenamefont {Liu}, \citenamefont {Zhu}, \citenamefont {Xu}, \citenamefont
  {Yang}, \citenamefont {Shang}, \citenamefont {Liu},\ and\ \citenamefont
  {Liao}}]{Ye2023}%
  \BibitemOpen
  \bibfield  {author} {\bibinfo {author} {\bibfnamefont {X.-G.}\ \bibnamefont
  {Ye}}, \bibinfo {author} {\bibfnamefont {H.}~\bibnamefont {Liu}}, \bibinfo
  {author} {\bibfnamefont {P.-F.}\ \bibnamefont {Zhu}}, \bibinfo {author}
  {\bibfnamefont {W.-Z.}\ \bibnamefont {Xu}}, \bibinfo {author} {\bibfnamefont
  {S.~A.}\ \bibnamefont {Yang}}, \bibinfo {author} {\bibfnamefont
  {N.}~\bibnamefont {Shang}}, \bibinfo {author} {\bibfnamefont
  {K.}~\bibnamefont {Liu}},\ and\ \bibinfo {author} {\bibfnamefont {Z.-M.}\
  \bibnamefont {Liao}},\ }\bibfield  {title} {\bibinfo {title} {Control over
  berry curvature dipole with electric field in {WTe}$_2$},\ }\href
  {https://doi.org/10.1103/physrevlett.130.016301} {\bibfield  {journal}
  {\bibinfo  {journal} {Phys. Rev. Lett.}\ }\textbf {\bibinfo {volume} {130}},\
  \bibinfo {pages} {016301} (\bibinfo {year} {2023})}\BibitemShut {NoStop}%
\bibitem [{\citenamefont {Wang}\ and\ \citenamefont {Qian}(2019)}]{Wang2019}%
  \BibitemOpen
  \bibfield  {author} {\bibinfo {author} {\bibfnamefont {H.}~\bibnamefont
  {Wang}}\ and\ \bibinfo {author} {\bibfnamefont {X.}~\bibnamefont {Qian}},\
  }\bibfield  {title} {\bibinfo {title} {Ferroelectric nonlinear anomalous hall
  effect in few-layer {WTe}$_2$},\ }\href
  {https://doi.org/10.1038/s41524-019-0257-1} {\bibfield  {journal} {\bibinfo
  {journal} {npj Comput. Mater.}\ }\textbf {\bibinfo {volume} {5}},\ \bibinfo
  {pages} {119} (\bibinfo {year} {2019})}\BibitemShut {NoStop}%
\bibitem [{\citenamefont {Battilomo}\ \emph {et~al.}(2019)\citenamefont
  {Battilomo}, \citenamefont {Scopigno},\ and\ \citenamefont
  {Ortix}}]{Battilomo2019}%
  \BibitemOpen
  \bibfield  {author} {\bibinfo {author} {\bibfnamefont {R.}~\bibnamefont
  {Battilomo}}, \bibinfo {author} {\bibfnamefont {N.}~\bibnamefont
  {Scopigno}},\ and\ \bibinfo {author} {\bibfnamefont {C.}~\bibnamefont
  {Ortix}},\ }\bibfield  {title} {\bibinfo {title} {Berry curvature dipole in
  strained graphene: A fermi surface warping effect},\ }\href
  {https://doi.org/10.1103/physrevlett.123.196403} {\bibfield  {journal}
  {\bibinfo  {journal} {Phys. Rev. Lett.}\ }\textbf {\bibinfo {volume} {123}},\
  \bibinfo {pages} {196403} (\bibinfo {year} {2019})}\BibitemShut {NoStop}%
\bibitem [{\citenamefont {Pantaleón}\ \emph {et~al.}(2021)\citenamefont
  {Pantaleón}, \citenamefont {Low},\ and\ \citenamefont
  {Guinea}}]{Pantaleon2021}%
  \BibitemOpen
  \bibfield  {author} {\bibinfo {author} {\bibfnamefont {P.~A.}\ \bibnamefont
  {Pantaleón}}, \bibinfo {author} {\bibfnamefont {T.}~\bibnamefont {Low}},\
  and\ \bibinfo {author} {\bibfnamefont {F.}~\bibnamefont {Guinea}},\
  }\bibfield  {title} {\bibinfo {title} {Tunable large berry dipole in strained
  twisted bilayer graphene},\ }\href
  {https://doi.org/10.1103/physrevb.103.205403} {\bibfield  {journal} {\bibinfo
   {journal} {Phys. Rev. B}\ }\textbf {\bibinfo {volume} {103}},\ \bibinfo
  {pages} {205403} (\bibinfo {year} {2021})}\BibitemShut {NoStop}%
\bibitem [{\citenamefont {Zhang}\ \emph {et~al.}(2022)\citenamefont {Zhang},
  \citenamefont {Xiao}, \citenamefont {Zhou}, \citenamefont {Hu}, \citenamefont
  {Xie}, \citenamefont {Yan},\ and\ \citenamefont {Law}}]{Zhang2022}%
  \BibitemOpen
  \bibfield  {author} {\bibinfo {author} {\bibfnamefont {C.-P.}\ \bibnamefont
  {Zhang}}, \bibinfo {author} {\bibfnamefont {J.}~\bibnamefont {Xiao}},
  \bibinfo {author} {\bibfnamefont {B.~T.}\ \bibnamefont {Zhou}}, \bibinfo
  {author} {\bibfnamefont {J.-X.}\ \bibnamefont {Hu}}, \bibinfo {author}
  {\bibfnamefont {Y.-M.}\ \bibnamefont {Xie}}, \bibinfo {author} {\bibfnamefont
  {B.}~\bibnamefont {Yan}},\ and\ \bibinfo {author} {\bibfnamefont {K.~T.}\
  \bibnamefont {Law}},\ }\bibfield  {title} {\bibinfo {title} {Giant nonlinear
  hall effect in strained twisted bilayer graphene},\ }\href
  {https://doi.org/10.1103/physrevb.106.l041111} {\bibfield  {journal}
  {\bibinfo  {journal} {Phys. Rev. B}\ }\textbf {\bibinfo {volume} {106}},\
  \bibinfo {pages} {l041111} (\bibinfo {year} {2022})}\BibitemShut {NoStop}%
\bibitem [{\citenamefont {Liu}\ \emph {et~al.}(2021)\citenamefont {Liu},
  \citenamefont {Zhao}, \citenamefont {Huang}, \citenamefont {Wu},
  \citenamefont {Sheng}, \citenamefont {Xiao},\ and\ \citenamefont
  {Yang}}]{Liu2021}%
  \BibitemOpen
  \bibfield  {author} {\bibinfo {author} {\bibfnamefont {H.}~\bibnamefont
  {Liu}}, \bibinfo {author} {\bibfnamefont {J.}~\bibnamefont {Zhao}}, \bibinfo
  {author} {\bibfnamefont {Y.-X.}\ \bibnamefont {Huang}}, \bibinfo {author}
  {\bibfnamefont {W.}~\bibnamefont {Wu}}, \bibinfo {author} {\bibfnamefont
  {X.-L.}\ \bibnamefont {Sheng}}, \bibinfo {author} {\bibfnamefont
  {C.}~\bibnamefont {Xiao}},\ and\ \bibinfo {author} {\bibfnamefont {S.~A.}\
  \bibnamefont {Yang}},\ }\bibfield  {title} {\bibinfo {title} {Intrinsic
  second-order anomalous hall effect and its application in compensated
  antiferromagnets},\ }\href {https://doi.org/10.1103/physrevlett.127.277202}
  {\bibfield  {journal} {\bibinfo  {journal} {Phys. Rev. Lett.}\ }\textbf
  {\bibinfo {volume} {127}},\ \bibinfo {pages} {277202} (\bibinfo {year}
  {2021})}\BibitemShut {NoStop}%
\bibitem [{\citenamefont {Wang}\ \emph
  {et~al.}(2023{\natexlab{a}})\citenamefont {Wang}, \citenamefont {Zeng},
  \citenamefont {Duan},\ and\ \citenamefont {Huang}}]{Wang2023}%
  \BibitemOpen
  \bibfield  {author} {\bibinfo {author} {\bibfnamefont {J.}~\bibnamefont
  {Wang}}, \bibinfo {author} {\bibfnamefont {H.}~\bibnamefont {Zeng}}, \bibinfo
  {author} {\bibfnamefont {W.}~\bibnamefont {Duan}},\ and\ \bibinfo {author}
  {\bibfnamefont {H.}~\bibnamefont {Huang}},\ }\bibfield  {title} {\bibinfo
  {title} {Intrinsic nonlinear hall detection of the néel vector for
  two-dimensional antiferromagnetic spintronics},\ }\href
  {https://doi.org/10.1103/physrevlett.131.056401} {\bibfield  {journal}
  {\bibinfo  {journal} {Phys. Rev. Lett.}\ }\textbf {\bibinfo {volume} {131}},\
  \bibinfo {pages} {056401} (\bibinfo {year} {2023}{\natexlab{a}})}\BibitemShut
  {NoStop}%
\bibitem [{\citenamefont {Gao}\ \emph {et~al.}(2023)\citenamefont {Gao},
  \citenamefont {Liu}, \citenamefont {Qiu}, \citenamefont {Ghosh},
  \citenamefont {V.~Trevisan}, \citenamefont {Onishi}, \citenamefont {Hu},
  \citenamefont {Qian}, \citenamefont {Tien}, \citenamefont {Chen},
  \citenamefont {Huang}, \citenamefont {Bérubé}, \citenamefont {Li},
  \citenamefont {Tzschaschel}, \citenamefont {Dinh}, \citenamefont {Sun},
  \citenamefont {Ho}, \citenamefont {Lien}, \citenamefont {Singh},
  \citenamefont {Watanabe}, \citenamefont {Taniguchi}, \citenamefont {Bell},
  \citenamefont {Lin}, \citenamefont {Chang}, \citenamefont {Du}, \citenamefont
  {Bansil}, \citenamefont {Fu}, \citenamefont {Ni}, \citenamefont {Orth},
  \citenamefont {Ma},\ and\ \citenamefont {Xu}}]{Gao2023}%
  \BibitemOpen
  \bibfield  {author} {\bibinfo {author} {\bibfnamefont {A.}~\bibnamefont
  {Gao}}, \bibinfo {author} {\bibfnamefont {Y.-F.}\ \bibnamefont {Liu}},
  \bibinfo {author} {\bibfnamefont {J.-X.}\ \bibnamefont {Qiu}}, \bibinfo
  {author} {\bibfnamefont {B.}~\bibnamefont {Ghosh}}, \bibinfo {author}
  {\bibfnamefont {T.}~\bibnamefont {V.~Trevisan}}, \bibinfo {author}
  {\bibfnamefont {Y.}~\bibnamefont {Onishi}}, \bibinfo {author} {\bibfnamefont
  {C.}~\bibnamefont {Hu}}, \bibinfo {author} {\bibfnamefont {T.}~\bibnamefont
  {Qian}}, \bibinfo {author} {\bibfnamefont {H.-J.}\ \bibnamefont {Tien}},
  \bibinfo {author} {\bibfnamefont {S.-W.}\ \bibnamefont {Chen}}, \bibinfo
  {author} {\bibfnamefont {M.}~\bibnamefont {Huang}}, \bibinfo {author}
  {\bibfnamefont {D.}~\bibnamefont {Bérubé}}, \bibinfo {author}
  {\bibfnamefont {H.}~\bibnamefont {Li}}, \bibinfo {author} {\bibfnamefont
  {C.}~\bibnamefont {Tzschaschel}}, \bibinfo {author} {\bibfnamefont
  {T.}~\bibnamefont {Dinh}}, \bibinfo {author} {\bibfnamefont {Z.}~\bibnamefont
  {Sun}}, \bibinfo {author} {\bibfnamefont {S.-C.}\ \bibnamefont {Ho}},
  \bibinfo {author} {\bibfnamefont {S.-W.}\ \bibnamefont {Lien}}, \bibinfo
  {author} {\bibfnamefont {B.}~\bibnamefont {Singh}}, \bibinfo {author}
  {\bibfnamefont {K.}~\bibnamefont {Watanabe}}, \bibinfo {author}
  {\bibfnamefont {T.}~\bibnamefont {Taniguchi}}, \bibinfo {author}
  {\bibfnamefont {D.~C.}\ \bibnamefont {Bell}}, \bibinfo {author}
  {\bibfnamefont {H.}~\bibnamefont {Lin}}, \bibinfo {author} {\bibfnamefont
  {T.-R.}\ \bibnamefont {Chang}}, \bibinfo {author} {\bibfnamefont {C.~R.}\
  \bibnamefont {Du}}, \bibinfo {author} {\bibfnamefont {A.}~\bibnamefont
  {Bansil}}, \bibinfo {author} {\bibfnamefont {L.}~\bibnamefont {Fu}}, \bibinfo
  {author} {\bibfnamefont {N.}~\bibnamefont {Ni}}, \bibinfo {author}
  {\bibfnamefont {P.~P.}\ \bibnamefont {Orth}}, \bibinfo {author}
  {\bibfnamefont {Q.}~\bibnamefont {Ma}},\ and\ \bibinfo {author}
  {\bibfnamefont {S.-Y.}\ \bibnamefont {Xu}},\ }\bibfield  {title} {\bibinfo
  {title} {Quantum metric nonlinear hall effect in a topological
  antiferromagnetic heterostructure},\ }\href
  {https://doi.org/10.1126/science.adf1506} {\bibfield  {journal} {\bibinfo
  {journal} {Science}\ }\textbf {\bibinfo {volume} {381}},\ \bibinfo {pages}
  {181} (\bibinfo {year} {2023})}\BibitemShut {NoStop}%
\bibitem [{\citenamefont {Wang}\ \emph
  {et~al.}(2023{\natexlab{b}})\citenamefont {Wang}, \citenamefont {Kaplan},
  \citenamefont {Zhang}, \citenamefont {Holder}, \citenamefont {Cao},
  \citenamefont {Wang}, \citenamefont {Zhou}, \citenamefont {Zhou},
  \citenamefont {Jiang}, \citenamefont {Zhang}, \citenamefont {Ru},
  \citenamefont {Cai}, \citenamefont {Watanabe}, \citenamefont {Taniguchi},
  \citenamefont {Yan},\ and\ \citenamefont {Gao}}]{Wang2023a}%
  \BibitemOpen
  \bibfield  {author} {\bibinfo {author} {\bibfnamefont {N.}~\bibnamefont
  {Wang}}, \bibinfo {author} {\bibfnamefont {D.}~\bibnamefont {Kaplan}},
  \bibinfo {author} {\bibfnamefont {Z.}~\bibnamefont {Zhang}}, \bibinfo
  {author} {\bibfnamefont {T.}~\bibnamefont {Holder}}, \bibinfo {author}
  {\bibfnamefont {N.}~\bibnamefont {Cao}}, \bibinfo {author} {\bibfnamefont
  {A.}~\bibnamefont {Wang}}, \bibinfo {author} {\bibfnamefont {X.}~\bibnamefont
  {Zhou}}, \bibinfo {author} {\bibfnamefont {F.}~\bibnamefont {Zhou}}, \bibinfo
  {author} {\bibfnamefont {Z.}~\bibnamefont {Jiang}}, \bibinfo {author}
  {\bibfnamefont {C.}~\bibnamefont {Zhang}}, \bibinfo {author} {\bibfnamefont
  {S.}~\bibnamefont {Ru}}, \bibinfo {author} {\bibfnamefont {H.}~\bibnamefont
  {Cai}}, \bibinfo {author} {\bibfnamefont {K.}~\bibnamefont {Watanabe}},
  \bibinfo {author} {\bibfnamefont {T.}~\bibnamefont {Taniguchi}}, \bibinfo
  {author} {\bibfnamefont {B.}~\bibnamefont {Yan}},\ and\ \bibinfo {author}
  {\bibfnamefont {W.}~\bibnamefont {Gao}},\ }\bibfield  {title} {\bibinfo
  {title} {Quantum-metric-induced nonlinear transport in a topological
  antiferromagnet},\ }\href {https://doi.org/10.1038/s41586-023-06363-3}
  {\bibfield  {journal} {\bibinfo  {journal} {Nature}\ }\textbf {\bibinfo
  {volume} {621}},\ \bibinfo {pages} {487} (\bibinfo {year}
  {2023}{\natexlab{b}})}\BibitemShut {NoStop}%
\bibitem [{\citenamefont {Gao}\ \emph {et~al.}(2014)\citenamefont {Gao},
  \citenamefont {Yang},\ and\ \citenamefont {Niu}}]{Gao2014}%
  \BibitemOpen
  \bibfield  {author} {\bibinfo {author} {\bibfnamefont {Y.}~\bibnamefont
  {Gao}}, \bibinfo {author} {\bibfnamefont {S.~A.}\ \bibnamefont {Yang}},\ and\
  \bibinfo {author} {\bibfnamefont {Q.}~\bibnamefont {Niu}},\ }\bibfield
  {title} {\bibinfo {title} {Field induced positional shift of bloch electrons
  and its dynamical implications},\ }\href
  {https://doi.org/10.1103/physrevlett.112.166601} {\bibfield  {journal}
  {\bibinfo  {journal} {Phys. Rev. Lett.}\ }\textbf {\bibinfo {volume} {112}},\
  \bibinfo {pages} {166601} (\bibinfo {year} {2014})}\BibitemShut {NoStop}%
\bibitem [{\citenamefont {Nandy}\ and\ \citenamefont
  {Sodemann}(2019)}]{Nandy2019}%
  \BibitemOpen
  \bibfield  {author} {\bibinfo {author} {\bibfnamefont {S.}~\bibnamefont
  {Nandy}}\ and\ \bibinfo {author} {\bibfnamefont {I.}~\bibnamefont
  {Sodemann}},\ }\bibfield  {title} {\bibinfo {title} {Symmetry and quantum
  kinetics of the nonlinear hall effect},\ }\href
  {https://doi.org/10.1103/physrevb.100.195117} {\bibfield  {journal} {\bibinfo
   {journal} {Phys. Rev. B}\ }\textbf {\bibinfo {volume} {100}},\ \bibinfo
  {pages} {195117} (\bibinfo {year} {2019})}\BibitemShut {NoStop}%
\bibitem [{\citenamefont {Das}\ \emph {et~al.}(2023)\citenamefont {Das},
  \citenamefont {Lahiri}, \citenamefont {Atencia}, \citenamefont {Culcer},\
  and\ \citenamefont {Agarwal}}]{Das2023}%
  \BibitemOpen
  \bibfield  {author} {\bibinfo {author} {\bibfnamefont {K.}~\bibnamefont
  {Das}}, \bibinfo {author} {\bibfnamefont {S.}~\bibnamefont {Lahiri}},
  \bibinfo {author} {\bibfnamefont {R.~B.}\ \bibnamefont {Atencia}}, \bibinfo
  {author} {\bibfnamefont {D.}~\bibnamefont {Culcer}},\ and\ \bibinfo {author}
  {\bibfnamefont {A.}~\bibnamefont {Agarwal}},\ }\bibfield  {title} {\bibinfo
  {title} {Intrinsic nonlinear conductivities induced by the quantum metric},\
  }\href {https://doi.org/10.1103/physrevb.108.l201405} {\bibfield  {journal}
  {\bibinfo  {journal} {Phys. Rev. B}\ }\textbf {\bibinfo {volume} {108}},\
  \bibinfo {pages} {l201405} (\bibinfo {year} {2023})}\BibitemShut {NoStop}%
\bibitem [{\citenamefont {Oiwa}\ and\ \citenamefont
  {Kusunose}(2022)}]{Oiwa2022}%
  \BibitemOpen
  \bibfield  {author} {\bibinfo {author} {\bibfnamefont {R.}~\bibnamefont
  {Oiwa}}\ and\ \bibinfo {author} {\bibfnamefont {H.}~\bibnamefont
  {Kusunose}},\ }\bibfield  {title} {\bibinfo {title} {Systematic analysis
  method for nonlinear response tensors},\ }\href
  {https://doi.org/10.7566/jpsj.91.014701} {\bibfield  {journal} {\bibinfo
  {journal} {J. Phys. Soc. Jpn.}\ }\textbf {\bibinfo {volume} {91}},\ \bibinfo
  {pages} {014701} (\bibinfo {year} {2022})}\BibitemShut {NoStop}%
\bibitem [{\citenamefont {Kaplan}\ \emph {et~al.}(2022)\citenamefont {Kaplan},
  \citenamefont {Holder},\ and\ \citenamefont {Yan}}]{Kaplan2022}%
  \BibitemOpen
  \bibfield  {author} {\bibinfo {author} {\bibfnamefont {D.}~\bibnamefont
  {Kaplan}}, \bibinfo {author} {\bibfnamefont {T.}~\bibnamefont {Holder}},\
  and\ \bibinfo {author} {\bibfnamefont {B.}~\bibnamefont {Yan}},\ }\bibfield
  {title} {\bibinfo {title} {Unification of nonlinear anomalous hall effect and
  nonreciprocal magnetoresistance in metals by the quantum geometry},\
  }\href@noop {} {\bibfield  {journal} {\bibinfo  {journal} {arXiv}\ }
  (\bibinfo {year} {2022})},\ \Eprint {https://arxiv.org/abs/2211.17213}
  {arXiv:2211.17213 [cond-mat.mes-hall]} \BibitemShut {NoStop}%
\bibitem [{\citenamefont {Michishita}\ and\ \citenamefont
  {Nagaosa}(2022)}]{Michishita2022}%
  \BibitemOpen
  \bibfield  {author} {\bibinfo {author} {\bibfnamefont {Y.}~\bibnamefont
  {Michishita}}\ and\ \bibinfo {author} {\bibfnamefont {N.}~\bibnamefont
  {Nagaosa}},\ }\bibfield  {title} {\bibinfo {title} {Dissipation and geometry
  in nonlinear quantum transports of multiband electronic systems},\ }\href
  {https://doi.org/10.1103/physrevb.106.125114} {\bibfield  {journal} {\bibinfo
   {journal} {Phys. Rev. B}\ }\textbf {\bibinfo {volume} {106}},\ \bibinfo
  {pages} {125114} (\bibinfo {year} {2022})}\BibitemShut {NoStop}%
\bibitem [{\citenamefont {Wang}\ \emph {et~al.}(2024)\citenamefont {Wang},
  \citenamefont {Zhang}, \citenamefont {Zhu},\ and\ \citenamefont
  {Su}}]{Wang2024}%
  \BibitemOpen
  \bibfield  {author} {\bibinfo {author} {\bibfnamefont {Y.}~\bibnamefont
  {Wang}}, \bibinfo {author} {\bibfnamefont {Z.}~\bibnamefont {Zhang}},
  \bibinfo {author} {\bibfnamefont {Z.-G.}\ \bibnamefont {Zhu}},\ and\ \bibinfo
  {author} {\bibfnamefont {G.}~\bibnamefont {Su}},\ }\bibfield  {title}
  {\bibinfo {title} {Intrinsic nonlinear ohmic current},\ }\href
  {https://doi.org/10.1103/physrevb.109.085419} {\bibfield  {journal} {\bibinfo
   {journal} {Phys. Rev. B}\ }\textbf {\bibinfo {volume} {109}},\ \bibinfo
  {pages} {085419} (\bibinfo {year} {2024})}\BibitemShut {NoStop}%
\bibitem [{\citenamefont {Zhang}\ \emph {et~al.}(2021)\citenamefont {Zhang},
  \citenamefont {Zhu},\ and\ \citenamefont {Su}}]{Zhang2021}%
  \BibitemOpen
  \bibfield  {author} {\bibinfo {author} {\bibfnamefont {Z.-F.}\ \bibnamefont
  {Zhang}}, \bibinfo {author} {\bibfnamefont {Z.-G.}\ \bibnamefont {Zhu}},\
  and\ \bibinfo {author} {\bibfnamefont {G.}~\bibnamefont {Su}},\ }\bibfield
  {title} {\bibinfo {title} {Theory of nonlinear response for charge and spin
  currents},\ }\href {https://doi.org/10.1103/physrevb.104.115140} {\bibfield
  {journal} {\bibinfo  {journal} {Phys. Rev. B}\ }\textbf {\bibinfo {volume}
  {104}},\ \bibinfo {pages} {115140} (\bibinfo {year} {2021})}\BibitemShut
  {NoStop}%
\bibitem [{\citenamefont {Shen}(2017)}]{Shen2017}%
  \BibitemOpen
  \bibfield  {author} {\bibinfo {author} {\bibfnamefont {S.~Q.}\ \bibnamefont
  {Shen}},\ }\href@noop {} {\emph {\bibinfo {title} {Topological
  Insulators-Dirac Equation in Condensed Matter, 2rd edition}}}\ (\bibinfo
  {publisher} {Springer, New York},\ \bibinfo {year} {2017})\BibitemShut
  {NoStop}%
\bibitem [{\citenamefont {Hamamoto}\ \emph {et~al.}(2017)\citenamefont
  {Hamamoto}, \citenamefont {Ezawa}, \citenamefont {Kim}, \citenamefont
  {Morimoto},\ and\ \citenamefont {Nagaosa}}]{Hamamoto2017}%
  \BibitemOpen
  \bibfield  {author} {\bibinfo {author} {\bibfnamefont {K.}~\bibnamefont
  {Hamamoto}}, \bibinfo {author} {\bibfnamefont {M.}~\bibnamefont {Ezawa}},
  \bibinfo {author} {\bibfnamefont {K.~W.}\ \bibnamefont {Kim}}, \bibinfo
  {author} {\bibfnamefont {T.}~\bibnamefont {Morimoto}},\ and\ \bibinfo
  {author} {\bibfnamefont {N.}~\bibnamefont {Nagaosa}},\ }\bibfield  {title}
  {\bibinfo {title} {Nonlinear spin current generation in noncentrosymmetric
  spin-orbit coupled systems},\ }\href
  {https://doi.org/10.1103/physrevb.95.224430} {\bibfield  {journal} {\bibinfo
  {journal} {Phys. Rev. B}\ }\textbf {\bibinfo {volume} {95}},\ \bibinfo
  {pages} {224430} (\bibinfo {year} {2017})}\BibitemShut {NoStop}%
\bibitem [{\citenamefont {Hayami}\ \emph {et~al.}(2022)\citenamefont {Hayami},
  \citenamefont {Yatsushiro},\ and\ \citenamefont {Kusunose}}]{Hayami2022}%
  \BibitemOpen
  \bibfield  {author} {\bibinfo {author} {\bibfnamefont {S.}~\bibnamefont
  {Hayami}}, \bibinfo {author} {\bibfnamefont {M.}~\bibnamefont {Yatsushiro}},\
  and\ \bibinfo {author} {\bibfnamefont {H.}~\bibnamefont {Kusunose}},\
  }\bibfield  {title} {\bibinfo {title} {Nonlinear spin hall effect in
  {PT}-symmetric collinear magnets},\ }\href
  {https://doi.org/10.1103/physrevb.106.024405} {\bibfield  {journal} {\bibinfo
   {journal} {Phys. Rev. B}\ }\textbf {\bibinfo {volume} {106}},\ \bibinfo
  {pages} {024405} (\bibinfo {year} {2022})}\BibitemShut {NoStop}%
\bibitem [{\citenamefont {Murakami}\ \emph
  {et~al.}(2004{\natexlab{a}})\citenamefont {Murakami}, \citenamefont
  {Nagaosa},\ and\ \citenamefont {Zhang}}]{Murakami2004}%
  \BibitemOpen
  \bibfield  {author} {\bibinfo {author} {\bibfnamefont {S.}~\bibnamefont
  {Murakami}}, \bibinfo {author} {\bibfnamefont {N.}~\bibnamefont {Nagaosa}},\
  and\ \bibinfo {author} {\bibfnamefont {S.-C.}\ \bibnamefont {Zhang}},\
  }\bibfield  {title} {\bibinfo {title} {Spin-hall insulator},\ }\href
  {https://doi.org/10.1103/physrevlett.93.156804} {\bibfield  {journal}
  {\bibinfo  {journal} {Phys. Rev. Lett.}\ }\textbf {\bibinfo {volume} {93}},\
  \bibinfo {pages} {156804} (\bibinfo {year} {2004}{\natexlab{a}})}\BibitemShut
  {NoStop}%
\bibitem [{\citenamefont {Murakami}\ \emph
  {et~al.}(2004{\natexlab{b}})\citenamefont {Murakami}, \citenamefont
  {Nagosa},\ and\ \citenamefont {Zhang}}]{Murakami2004a}%
  \BibitemOpen
  \bibfield  {author} {\bibinfo {author} {\bibfnamefont {S.}~\bibnamefont
  {Murakami}}, \bibinfo {author} {\bibfnamefont {N.}~\bibnamefont {Nagosa}},\
  and\ \bibinfo {author} {\bibfnamefont {S.-C.}\ \bibnamefont {Zhang}},\
  }\bibfield  {title} {\bibinfo {title} {Su$(2)$ non-abelian holonomy and
  dissipationless spin current in semiconductors},\ }\href
  {https://doi.org/10.1103/physrevb.69.235206} {\bibfield  {journal} {\bibinfo
  {journal} {Phys. Rev. B}\ }\textbf {\bibinfo {volume} {69}},\ \bibinfo
  {pages} {235206} (\bibinfo {year} {2004}{\natexlab{b}})}\BibitemShut
  {NoStop}%
\bibitem [{\citenamefont {Sinova}\ \emph {et~al.}(2004)\citenamefont {Sinova},
  \citenamefont {Culcer}, \citenamefont {Niu}, \citenamefont {Sinitsyn},
  \citenamefont {Jungwirth},\ and\ \citenamefont {MacDonald}}]{Sinova2004}%
  \BibitemOpen
  \bibfield  {author} {\bibinfo {author} {\bibfnamefont {J.}~\bibnamefont
  {Sinova}}, \bibinfo {author} {\bibfnamefont {D.}~\bibnamefont {Culcer}},
  \bibinfo {author} {\bibfnamefont {Q.}~\bibnamefont {Niu}}, \bibinfo {author}
  {\bibfnamefont {N.~A.}\ \bibnamefont {Sinitsyn}}, \bibinfo {author}
  {\bibfnamefont {T.}~\bibnamefont {Jungwirth}},\ and\ \bibinfo {author}
  {\bibfnamefont {A.~H.}\ \bibnamefont {MacDonald}},\ }\bibfield  {title}
  {\bibinfo {title} {Universal intrinsic spin hall effect},\ }\href
  {https://doi.org/10.1103/physrevlett.92.126603} {\bibfield  {journal}
  {\bibinfo  {journal} {Phys. Rev. Lett.}\ }\textbf {\bibinfo {volume} {92}},\
  \bibinfo {pages} {126603} (\bibinfo {year} {2004})}\BibitemShut {NoStop}%
\bibitem [{\citenamefont {Gao}(2019)}]{Gao2019}%
  \BibitemOpen
  \bibfield  {author} {\bibinfo {author} {\bibfnamefont {Y.}~\bibnamefont
  {Gao}},\ }\bibfield  {title} {\bibinfo {title} {Semiclassical dynamics and
  nonlinear charge current},\ }\href
  {https://doi.org/10.1007/s11467-019-0887-2} {\bibfield  {journal} {\bibinfo
  {journal} {Front. Phys.}\ }\textbf {\bibinfo {volume} {14(3)}},\ \bibinfo
  {pages} {33404} (\bibinfo {year} {2019})}\BibitemShut {NoStop}%
\bibitem [{\citenamefont {Yao}\ \emph {et~al.}(2004)\citenamefont {Yao},
  \citenamefont {Kleinman}, \citenamefont {MacDonald}, \citenamefont {Sinova},
  \citenamefont {Jungwirth}, \citenamefont {Wang}, \citenamefont {Wang},\ and\
  \citenamefont {Niu}}]{Yao2004}%
  \BibitemOpen
  \bibfield  {author} {\bibinfo {author} {\bibfnamefont {Y.}~\bibnamefont
  {Yao}}, \bibinfo {author} {\bibfnamefont {L.}~\bibnamefont {Kleinman}},
  \bibinfo {author} {\bibfnamefont {A.~H.}\ \bibnamefont {MacDonald}}, \bibinfo
  {author} {\bibfnamefont {J.}~\bibnamefont {Sinova}}, \bibinfo {author}
  {\bibfnamefont {T.}~\bibnamefont {Jungwirth}}, \bibinfo {author}
  {\bibfnamefont {D.-s.}\ \bibnamefont {Wang}}, \bibinfo {author}
  {\bibfnamefont {E.}~\bibnamefont {Wang}},\ and\ \bibinfo {author}
  {\bibfnamefont {Q.}~\bibnamefont {Niu}},\ }\bibfield  {title} {\bibinfo
  {title} {First principles calculation of anomalous hall conductivity in
  ferromagnetic bcc {Fe}},\ }\href
  {https://doi.org/10.1103/physrevlett.92.037204} {\bibfield  {journal}
  {\bibinfo  {journal} {Phys. Rev. Lett.}\ }\textbf {\bibinfo {volume} {92}},\
  \bibinfo {pages} {037204} (\bibinfo {year} {2004})}\BibitemShut {NoStop}%
\bibitem [{\citenamefont {Guo}\ \emph {et~al.}(2005)\citenamefont {Guo},
  \citenamefont {Yao},\ and\ \citenamefont {Niu}}]{Guo2005}%
  \BibitemOpen
  \bibfield  {author} {\bibinfo {author} {\bibfnamefont {G.~Y.}\ \bibnamefont
  {Guo}}, \bibinfo {author} {\bibfnamefont {Y.}~\bibnamefont {Yao}},\ and\
  \bibinfo {author} {\bibfnamefont {Q.}~\bibnamefont {Niu}},\ }\bibfield
  {title} {\bibinfo {title} {Ab initio calculation of the intrinsic spin hall
  effect in semiconductors},\ }\href
  {https://doi.org/10.1103/physrevlett.94.226601} {\bibfield  {journal}
  {\bibinfo  {journal} {Phys. Rev. Lett.}\ }\textbf {\bibinfo {volume} {94}},\
  \bibinfo {pages} {226601} (\bibinfo {year} {2005})}\BibitemShut {NoStop}%
\bibitem [{\citenamefont {Guo}\ \emph {et~al.}(2008)\citenamefont {Guo},
  \citenamefont {Murakami}, \citenamefont {Chen},\ and\ \citenamefont
  {Nagaosa}}]{Guo2008}%
  \BibitemOpen
  \bibfield  {author} {\bibinfo {author} {\bibfnamefont {G.~Y.}\ \bibnamefont
  {Guo}}, \bibinfo {author} {\bibfnamefont {S.}~\bibnamefont {Murakami}},
  \bibinfo {author} {\bibfnamefont {T.-W.}\ \bibnamefont {Chen}},\ and\
  \bibinfo {author} {\bibfnamefont {N.}~\bibnamefont {Nagaosa}},\ }\bibfield
  {title} {\bibinfo {title} {Intrinsic spin hall effect in platinum:
  First-principles calculations},\ }\href
  {https://doi.org/10.1103/physrevlett.100.096401} {\bibfield  {journal}
  {\bibinfo  {journal} {Phys. Rev. Lett.}\ }\textbf {\bibinfo {volume} {100}},\
  \bibinfo {pages} {096401} (\bibinfo {year} {2008})}\BibitemShut {NoStop}%
\bibitem [{\citenamefont {Chen}(2023)}]{Chen2023}%
  \BibitemOpen
  \bibfield  {author} {\bibinfo {author} {\bibfnamefont {W.}~\bibnamefont
  {Chen}},\ }\bibfield  {title} {\bibinfo {title} {Optical absorption
  measurement of spin berry curvature and spin chern marker},\ }\href{https://doi.org/10.1088/1361-648X/acba72}
  {\bibfield  {journal} {\bibinfo  {journal} {J. Phys. Condens. Matter}\
  }\textbf {\bibinfo {volume} {35}},\ \bibinfo {pages} {155601} (\bibinfo
  {year} {2023})}\BibitemShut {NoStop}%
\bibitem [{\citenamefont {Mahan}(2000)}]{Mahan}%
  \BibitemOpen
  \bibfield  {author} {\bibinfo {author} {\bibfnamefont {G.~D.}\ \bibnamefont
  {Mahan}},\ }\href@noop {} {\emph {\bibinfo {title} {Many-particle physics,
  3rd edition}}}\ (\bibinfo  {publisher} {Kluwer Acdemic Publishers, New
  York},\ \bibinfo {year} {2000})\BibitemShut {NoStop}%
\bibitem [{\citenamefont {Rashba}(2003)}]{Rashba2003}%
  \BibitemOpen
  \bibfield  {author} {\bibinfo {author} {\bibfnamefont {E.~I.}\ \bibnamefont
  {Rashba}},\ }\bibfield  {title} {\bibinfo {title} {Spin currents in
  thermodynamic equilibrium: The challenge of discerning transport currents},\
  }\href {https://doi.org/10.1103/physrevb.68.241315} {\bibfield  {journal}
  {\bibinfo  {journal} {Phys. Rev. B}\ }\textbf {\bibinfo {volume} {68}},\
  \bibinfo {pages} {241315} (\bibinfo {year} {2003})}\BibitemShut {NoStop}%
\bibitem [{\citenamefont {Feng}\ \emph {et~al.}(2012)\citenamefont {Feng},
  \citenamefont {Yao}, \citenamefont {Zhu}, \citenamefont {Zhou}, \citenamefont
  {Yao},\ and\ \citenamefont {Xiao}}]{Feng2012}%
  \BibitemOpen
  \bibfield  {author} {\bibinfo {author} {\bibfnamefont {W.}~\bibnamefont
  {Feng}}, \bibinfo {author} {\bibfnamefont {Y.}~\bibnamefont {Yao}}, \bibinfo
  {author} {\bibfnamefont {W.}~\bibnamefont {Zhu}}, \bibinfo {author}
  {\bibfnamefont {J.}~\bibnamefont {Zhou}}, \bibinfo {author} {\bibfnamefont
  {W.}~\bibnamefont {Yao}},\ and\ \bibinfo {author} {\bibfnamefont
  {D.}~\bibnamefont {Xiao}},\ }\bibfield  {title} {\bibinfo {title} {Intrinsic
  spin hall effect in monolayers of group-vi dichalcogenides: A
  first-principles study},\ }\href {https://doi.org/10.1103/physrevb.86.165108}
  {\bibfield  {journal} {\bibinfo  {journal} {Phys. Rev. B}\ }\textbf {\bibinfo
  {volume} {86}},\ \bibinfo {pages} {165108} (\bibinfo {year}
  {2012})}\BibitemShut {NoStop}%
\bibitem [{\citenamefont {Kapri}\ \emph {et~al.}(2021)\citenamefont {Kapri},
  \citenamefont {Dey},\ and\ \citenamefont {Ghosh}}]{Kapri2021}%
  \BibitemOpen
  \bibfield  {author} {\bibinfo {author} {\bibfnamefont {P.}~\bibnamefont
  {Kapri}}, \bibinfo {author} {\bibfnamefont {B.}~\bibnamefont {Dey}},\ and\
  \bibinfo {author} {\bibfnamefont {T.~K.}\ \bibnamefont {Ghosh}},\ }\bibfield
  {title} {\bibinfo {title} {Role of berry curvature in the generation of spin
  currents in rashba systems},\ }\href
  {https://doi.org/10.1103/physrevb.103.165401} {\bibfield  {journal} {\bibinfo
   {journal} {Phys. Rev. B}\ }\textbf {\bibinfo {volume} {103}},\ \bibinfo
  {pages} {165401} (\bibinfo {year} {2021})}\BibitemShut {NoStop}%
\bibitem [{\citenamefont {Muñoz-Santana}\ and\ \citenamefont
  {Maytorena}(2023)}]{MunozSantana2023}%
  \BibitemOpen
  \bibfield  {author} {\bibinfo {author} {\bibfnamefont {D.}~\bibnamefont
  {Muñoz-Santana}}\ and\ \bibinfo {author} {\bibfnamefont {J.~A.}\
  \bibnamefont {Maytorena}},\ }\bibfield  {title} {\bibinfo {title} {Linear and
  nonlinear spin current response of anisotropic spin-orbit coupled systems},\
  }\href {https://doi.org/10.1088/1361-648x/acf74d} {\bibfield  {journal}
  {\bibinfo  {journal} {J. Phys.: Condens. Matter}\ }\textbf {\bibinfo {volume}
  {35}},\ \bibinfo {pages} {505301} (\bibinfo {year} {2023})}\BibitemShut
  {NoStop}%
\bibitem [{\citenamefont {Lihm}\ and\ \citenamefont {Park}(2022)}]{Lihm2022}%
  \BibitemOpen
  \bibfield  {author} {\bibinfo {author} {\bibfnamefont {J.-M.}\ \bibnamefont
  {Lihm}}\ and\ \bibinfo {author} {\bibfnamefont {C.-H.}\ \bibnamefont
  {Park}},\ }\bibfield  {title} {\bibinfo {title} {Comprehensive theory of
  second-order spin photocurrents},\ }\href
  {https://doi.org/10.1103/PhysRevB.105.045201} {\bibfield  {journal} {\bibinfo
   {journal} {Phys. Rev. B}\ }\textbf {\bibinfo {volume} {105}},\ \bibinfo
  {pages} {045201} (\bibinfo {year} {2022})}\BibitemShut {NoStop}
\end{thebibliography}


\end{document}